\documentclass[11pt]{article}

\usepackage[normalem]{ulem}
\usepackage{amsmath, amssymb,amsthm,bm,bbm,fullpage}
\usepackage{mathrsfs}
\usepackage{cite}
\usepackage{colortbl}
\usepackage[all]{xy}
\usepackage{epsfig}
\usepackage{subfigure}
\usepackage{graphics}
\usepackage{multirow}
\usepackage{lineno}
\usepackage{setspace}
\usepackage{caption}
\usepackage[colorlinks=true, allcolors=black]{hyperref}

\usepackage{graphicx}

\theoremstyle{remark} 
	
	\doublespace

\begin{document}

\begin{centering}
{\huge
\textbf{Cultural tightness and social cohesion under evolving norms}
}
\bigskip
\\
Filippo Zimmaro\footnote{Department of Mathematics, University of Bologna, Italy,  Department of Computer Science, University of Pisa, Italy}, Jacopo Grilli\footnote{The Abdus Salam International Centre for Theoretical Physics (ICTP), Strada Costiera 11, 34014 Trieste, Italy}, Mirta Galesic\footnote{Complexity Science Hub Vienna, Austria, Santa Fe Institute, Santa Fe, NM, USA, Vermont Complex Systems Institute, University of Vermont, Burlington, VT, USA}, Alexander J. Stewart\footnote{Center for Complex Networks and Systems Research, Luddy School of Informatics, Computing, and Engineering, Indiana University, Bloomington, IN, USA}\footnote{E-mail: stewalex@iu.edu}
\\
\bigskip
\end{centering}

\noindent\textbf{Successful collective action on issues from climate change to the maintenance of democracy depends on societal properties such as cultural tightness and social cohesion. How these properties evolve is not well understood because they emerge from a complex interplay between beliefs and behaviors that are usually modeled separately. Here we address this challenge by developing a game-theoretical framework incorporating norm-utility models to study the coevolutionary dynamics of cooperative action, expressed belief, and norm-utility preferences. We show that the introduction of evolving beliefs and preferences into the Snowdrift game and Prisoner's Dilemma leads to a proliferation of evolutionary stable equilibria, each with different societal properties. In particular, we find that a declining material environment can simultaneously be associated with increased cultural tightness (defined as the degree to which individuals behave in accordance with widely held beliefs) and reduced social cohesion (defined as the degree of social homogeneity i.e. the extent to which individuals belong to a single well-defined group). Loss of social homogeneity occurs via a process of evolutionary branching, in which a population fragments into two distinct social groups with strikingly different characteristics. The groups that emerge differ not only in their willingness to cooperate, but also in their beliefs about cooperation and in their preferences for conformity and coherence of their actions and beliefs. These results have implications for our understanding of the resilience of cooperation and collective action in times of crisis. }

\clearpage

\noindent People’s behaviors in social settings are not always in line with their expressed beliefs. They support political candidates whose policies do not align with their professed values \cite{converse2006nature,lau1997voting}, and treat others in ways that appear to violate the moral norms they claim to follow \cite{gerlach2019truth,hallman2023theory}. Such discrepancies also occur on institutional levels. For example, despite a lot of political posturing, the UN fund for damages due to climate change has so far received only around $700$ million dollars worth of pledges  \cite{unfccc_loss_damage_pledges_2025}, a small fraction of more than $500$ billion dollars needed to cover the damages experienced by the most vulnerable countries \cite{v20_loss_report_2022}.

The way people align their beliefs and behaviors in social settings depends not only on the material payoffs they derive from their actions, but also on psychological utility derived from adherence to social norms, i.e. norm utility \cite{weber2019economy, boudon2009rationalite}. People can have different preferences related to norm utility, including the extent to which they prefer psychological vs. material payoffs, and the relative preference for \emph{coherence} -- i.e. for aligning one's own belief and behavior --  vs. for \emph{conformity} -- i.e. for aligning one's belief to other people's beliefs (\emph{injunctive norm}) or to other people's behavior (\emph{descriptive norm}) \cite{gavrilets2024modelling,heinicke2022injunctive}. Descriptive norms can facilitate the alignment of behaviors and beliefs, but may be hard to discern when there is a lack of information or a lot of behavioral variation \cite{morris2015normology}. In contrast, injunctive norms, which can be inferred through communication \cite{bicchieri2002covenants}, run the risk of inducing a shift in expressed beliefs that is not accompanied by a consistent behavior.

The dynamics of the interplay between beliefs, actions, and preferences can lead to societies with different higher-level properties. 
One such societal property is \emph{cultural tightness}, the degree to which individuals behave in line with societal norms 
\cite{gelfand2011differences,harrington2014tightness}. Differences in cultural tightness across societies has been linked to experiences with various threats, but their evolution remains insufficiently understood \cite{jackson2020global}. For example, while cultural tightness is generally expected to increase over time because all societies eventually experience threats, many tight cultures actually become looser over time and others fluctuate, as reflected in changing indicators of their democratic freedom \cite{gelfand2011differences,vdem_institute_2025}. It also remains unexplained why, within the same society, some groups exhibit lower and some higher levels of tightness (e.g. older men vs. younger women) \cite{jiang_societies_2015,qin_gender_2023}, and why tightness can be higher for some social norms than for others (e.g. hand washing norms during the COVID-19 pandemic, or norms about socialization, marriage, and mourning in some non-industrial societies)  \cite{andrighetto_changes_2024,jackson2020global}.  

Another such property is social cohesion, which has been defined in various ways, including prosocial behavior, social capital, and sense of belonging \cite{friedkin2004social,schiefer2017essentials}. One important aspect of social cohesion is social homogeneity, the extent to which individuals belong to a single, dominant group, as opposed to being distributed across multiple, fragmented groups. This may include groups who differ not only in their behavior, but also in their beliefs, and in their preferences for conformity or coherence. For example, some groups might believe in the value of cooperation, value social conformity and behave cooperatively, while others might be skeptical about cooperation, value intellectual coherence and defect. 

Despite its importance, the coevolution of beliefs, behaviors, and preferences has been underexplored theoretically. One reason for this is an historical disconnect between the literatures on belief dynamics, collective action, and the evolution of social norms. Models of belief dynamics \cite{castellano2009statistical,flache2017models,galesic2021integrating} typically do not investigate how beliefs relate to people’s behaviors, implicitly assuming that people will act in line with their beliefs. Models of collective action often assume that people have relatively fixed beliefs about societally relevant issues such as fairness, group identity, injustice, or the value and efficacy of collective action \cite{Fehr1999,ostrom2000collective,van_zomeren_toward_2008}. These models typically do not include mechanisms for belief change. Recent work has explored the joint dynamics of beliefs and behaviors \cite{bisin_economics_2001,calabuig_culture_2018,gavrilets_coevolution_2021,kuran_cultural_2008,tverskoi2024cultural,doi:10.1073/pnas.2504339122,Wu2024} but has not explicitly addressed either how emergent social properties such as cultural tightness evolve, nor considered the evolution of preferences for coherence and conformity of belief. 

Here we address this challenge in the context of cooperation. We develop a mathematical model describing the coevolution of cooperative action, belief in the value of cooperation, and preferences for coherence and conformity to descriptive and injunctive norms. We study the societal equilibria that evolve, characterized by the distribution of cooperation rates, beliefs and preferences across individuals. We then interpret these equilibria in terms of the level of cultural tightness and social homogeneity at the population level.


A distinctive feature of our model is that the relative importance of the preferences for coherence and conformity is not exogenously given, but endogenously determined and updated over time through social learning, in parallel with belief and behaviour. 
This evolution of preferences is seldom addressed in existing models (although see \cite{Ingela2023}), where such parameters are usually held fixed \cite{rabin1994cognitive, gavrilets2024modelling}. We show that when preferences are allowed to coevolve with beliefs and behaviors the number and type of social equilibria proliferate, with profound consequences for the emergent cultural tightness and the rate of cooperation in the population. Moreover, we find that the evolution of preferences reduces social cohesion through a process of evolutionary branching, in which the population of initially identical individuals splits into different groups, characterized by different behaviours, different beliefs and different preferences for coherence vs conformity.  The nature of the equilibria that evolve depends on the material costs and benefits of cooperation, and on how preferences are allowed to evolve. Most strikingly we show that there is a tension, such that a shift towards a more difficult material environment (e.g. higher costs of cooperation) can increase cultural tightness, but at the expense of social homogeneity.

\begin{figure}[tbhp]
\centering
\includegraphics[width=1.0\linewidth]{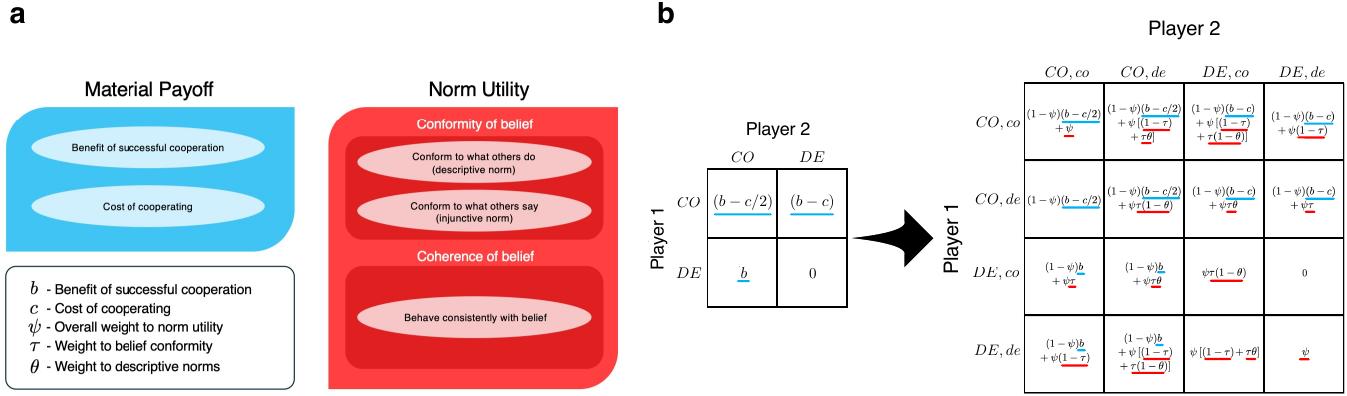}
\caption{\footnotesize \textbf{Incorporating belief into a normal form game}. a) We follow the norm utility approach \cite{gavrilets2024modelling} under which we consider a utility function that combines contributions from material payoff (blue) with contributions from adherence to perceived social norms (red). Material payoff is derived from the costs and benefits of cooperation ($b$ and $c$), while norm utility is derived from weighted contributions from belief conformity -- the extent to which an individual's expressed belief aligns with a perceived descriptive or injunctive norm -- and belief coherence -- alignment between expressed belief and behavior. The parameters for the different payoff contributions are shown in the box on the bottom left. b) In order to illustrate how  norm utility is incorporated into a normal-form game we transform the standard (material) payoff matrix (Left) for the Snowdrift game without belief (payoffs shown are for Player 1) into a matrix incorporating composite actions (right) of the form $X_ix_i$, where $X_i\in\{CO,DE\}$ is the material decision of player $i$ and $x_i\in\{co,de\}$
is their expressed belief. The resulting payoffs are a weighted sum of the material terms (underlined in blue) and contributions from psychological sources of utility (underlined in red), with components derived from conformity and coherence of belief as described in detail in the main text.}
\label{fig: the model}
\end{figure}

\section*{Model \& Results}

\noindent \textbf{Modelling framework.} 
We study the coevolution of behaviors, beliefs, and preferences in a game theoretic setting. The effects of beliefs can be integrated into game theoretic models in a number of ways. Our approach is to integrate belief \emph{expression} and norm utility directly into the payoff matrix of a normal form game. This is achieved by expanding the action space of the original game (Figure 1).
In the expanded action space, the players choose a composite action, comprised of a 
\emph{material behavior} 
 -- what they choose to physically do in the game, such as cooperate or defect -- and an \emph{expressed belief} -- what they choose to publicly state about what is the right thing to do. The material behavior of the players generates a material component of utility, determined by the costs and benefits of cooperation. The combination of material action and expressed belief generates a norm utility. In particular, both the preferences of the individual for conformity with others' beliefs and for coherence between their own material action and their expressed beliefs contribute to the norm utility (Figure \ref{fig: the model}). 

We take as our starting point a classic two-player game, in which the players simultaneously choose whether to cooperate in order to help solve a problem which requires cooperation and/or coordination (e.g. to clear a snowdrift that blocks a shared driveway). The total cost to solve the problem is $c$, and the benefit to each player if the problem is solved is $b$. If both players cooperate, they share the cost equally, whereas if only one player cooperates, they shoulder all of the cost on their own.  If both players defect, the problem remains unsolved, and neither receives any benefit or pays any cost. The case $c > 2 b$ (the sum of benefits of cooperation is always lower than the sum of its cost) is trivial and therefore we constrain the payoffs so that $0<c<2b$. When $0<c<b$ the game corresponds to a Snowdrift game, with a mixed Nash equilibrium, and when $b<c<2b$ the game is a Prisoner's Dilemma, with defect as the sole Nash equilibrium for the one-shot, two-players game.

In the standard game without any belief, each player simply chooses whether to \emph{cooperate} ($CO$) or \emph{defect} ($DE$). In the expanded action space incorporating the expression of belief, each player has four possible composite actions as follows: 
\begin{enumerate}
\item Cooperate and express a personal belief in cooperation ($CO,co$)
\item Cooperate and express a personal belief in defection ($CO,de$)
\item Defect and express a personal belief in cooperation ($DE,co$)
\item Defect and express a personal belief in defection ($DE,de$)
\end{enumerate}

We employ an expanded payoff matrix \cite{Wu2024} which captures both material payoff and norm utility (Figure \ref{fig: the model}b).
In our framework, norm utility arises from two sources: preference for \emph{conformity} of own belief to a norm, and preference for \emph{coherence} between one's expressed belief and material behavior (Figure 1). Two parameters --- $\psi$ and $\tau$ --- quantify the contributions of norm utility to overall utility. The norm-utility weight $\psi\in[0,1]$ describes the contribution of the norm utility to overall utility compared to the contribution of material factors (the costs and benefits of cooperation). When $\phi = 0$ the extended payoff reduces to the material one, while in the case $\psi = 1$ it fully corresponds to the norm utility. The conformity weight $\tau\in[0,1]$ defines the norm utility by quantifying the relative contribution of a preference for conformity as opposed to coherence (Figure 1). If $\tau = 0$ only coherence contribute to norm-utility, while if $\tau=1$ only conformity matters.

The conformity to a norm to others can occur in multiple way. In fact, individuals may conform by aligning their expressed belief with the expressed beliefs of others (injunctive norms). On the other extreme, they could align with the observed behavior of others (descriptive norms). We weight the importance of descriptive norms introducing a parameter $\theta\in[0,1]$. If $\theta = 0$ only the expressed beliefs contribute to utility, while if $\theta = 1$ utility is determined by descriptive norms.

We can encode material action of a player $i$ as a binary variable $X_i\in\{CO,DE\}$ and their expressed belief as $x_i\in\{co,de\}$. The overall utility $\Pi_{ij}$ from a given interaction between player $i$ and player $j$ is given by

\begin{align}
 \nonumber   &\Pi_{ij}(X_i,x_i,X_j,x_j)=\\
    &(1-\psi_i)\Pi_{ij}^\text{mat}(X_i,X_j)+\psi_i\Pi_{ij}^{\text{norm}}(X_i,x_i,X_j,x_j)
\end{align}

where $\Pi_{ij}^\text{mat}$ is the material payoff received by player $i$ from interacting with player $j$ (i.e. the payoff received from the standard normal form game without any belief) and $\Pi_{ij}^{\text{norm}}$ is the norm utility which depends on both actions and beliefs
\begin{align}
\nonumber&\Pi_{ij}^{\text{norm}}(X_i,x_i,X_j,x_j)=\\
&(1-\tau_i)\delta_{x_iX_i}+\tau_i\left(\theta_i\delta_{x_iX_j}+(1-\theta_i)\delta_{x_ix_j}\right)
\end{align}

 Here $\delta_{kl}$ is the Kronecker delta (which should be interpreted in a broad sense when comparing actions and beliefs, i.e., $\delta_{co,CO}=1$, see Methods). The maximum norm utility that can be achieved is $\Pi_{ij}^{\text{norm}}=1$, while the maximum material payoff is $\Pi_{ij}^\text{mat}=b$. In general, we expect the ratio $b/c$ determines the equilibria, as in standard games, but also  the overall magnitude of the benefit of cooperation $b$ to contribute.
\\
\\
\noindent \textbf{Evolution through social learning.}
The expanded game, with composite actions and utility derived from both material payoffs and norm utility, can be analyzed using standard approaches.
Here we focus on the co-evolution of behavior, belief and preferences. In particular we assume that not only actions and beliefs evolve but also the preference for conformity and coherence (quantified by $\psi$, $\theta$, and $\tau$) change over time under a model of imitation through social learning in a well-mixed population.

We consider a well-mixed population of $N$ players participating in a pairwise game with belief, as described above (Figure 1b). When players interact they simultaneously choose a material action (cooperate or defect) and also express a belief about the right decision. The action and the expression of belief are chosen probabilistically: $p_i$ is the probability that agent $i$ acts cooperatively ($1-p_i$ that it defects) and $q_i$ the probability that expresses a belief in cooperation. The payoff a player $i$ receives from such an interaction with an opponent $j$ is as given by Eq. 1, while the expected payoff over mixed strategies, $\pi_{i}$, is given by Eq. 11 (see SI). Since the population is well-mixed the expected payoff for a player $i$ is simply
$\pi_i=\frac{1}{N-1}\sum_{j\neq i}\pi_{ij}$. 

Evolution occurs through payoff-based imitation, a form of social learning, in which a player $k$ copies the strategy of a player $l$ with probability $\frac{1}{1+\exp[s(\pi_k-\pi_l)]}$, where $s$ is the intensity of selection \cite{Traulsen:2006zr}. In general, the ``strategy'' of a player, which gets imitated, comprises both their material behavior (the probability $p_i$  at which they cooperate or defect), their belief in cooperation (quantified by $q_i$), and their preferences for conformity and coherence (determined by $\psi$, $\tau$, and $\theta$). 

We first present results in which only belief and behavior are allowed to evolve, while keeping preferences fixed (i.e., $\tau_i=\tau$, $\psi_i=\psi$ and $\theta_i=\theta$ for all $i$). This case provides a baseline against which to compare the outcome of evolution when preferences are also allowed to evolve. Next, we present results when either the conformity preference $\tau$ or the overall norm-utility weight $\psi$ is allowed to evolve alongside belief and behavior (keeping the other parameters fixed). Finally, we characterize the evolutionary dynamics of the system when all three preference parameters ($\psi$, $\tau$, and $\theta$) evolve alongside belief and behavior.

Our results are presented in two forms: the results of invasion analysis, which characterize the equilibria of the evolutionary dynamics, and the results of individual-based simulations under the imitation process described above, which allow us to determine the basins of attraction associated with the different equilibria. Further details on the invasion analysis can be found in the SI section 2, and details of the individual-based simulations can be found in the Methods and SI section 3.
\\
\\
\noindent \textbf{Co-evolution of belief and behavior.} 
The probability $p_i$ that player $i$ cooperates in a given interaction, and $q_i$  that player $i$ expresses a belief in cooperation during the interaction are the traits subject to co-evolution.
Under this model, the probability that player $i$ chooses composite action $\{CO,co\}$ in a given interaction is $p_iq_i$, the probability they play $\{DE,de\}$ is $(1-p_i)(1-q_i)$ and so on. We use invasion analysis \cite{Mullon:2016aa} to study the evolutionary dynamics of $p$ and $q$ across the population. We compare the utility of an ``invading'' player $i$, with traits $(p_i,q_i)$, to the utility of the ``resident'' traits $(p_r,q_r)$, which are shared by all other members of the population. A pair of resident traits $(p_r^*,q_r^*)$ are evolutionary stable if they cannot be invaded by any available local mutant, i.e. by a strategy which is displaced from the resident by a small amount (see SI Section 1-2 for a detailed analysis of the Nash equilibria and the evolutionary stability of this system).

\begin{figure}[tbhp]
\centering
\includegraphics[width=0.75\linewidth]{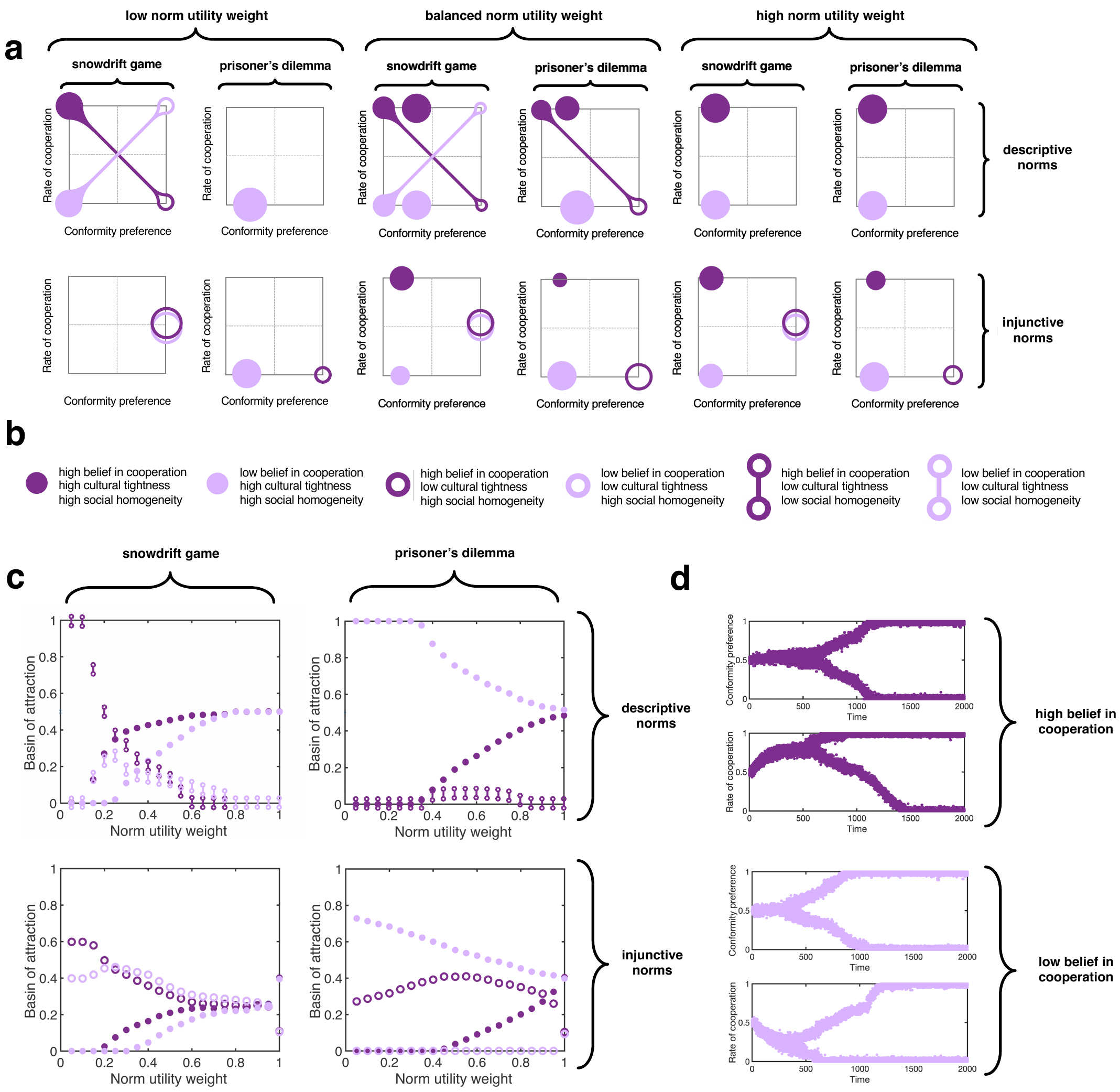}
\caption{\footnotesize \textbf{Co-evolution of cooperation, belief and conformity preference.} We performed individual-based simulations for the co-evolution of cooperation, $p$, belief in cooperation, $q$ and the preference for conformity as opposed to coherence, $\tau$, under injunctive norms ($\theta=0$), and descriptive norms ($\theta=1$). a) Qualitative characteristics of the different types of equilibria. We identify six possible equilibria for the system  which differ in the level of social homogeneity of the population (i.e. whether the population is monomorphic or polymorphic, see SI section 2 for a detailed description of the categorization of equilibria), their level of tightness (see Eq 3) and the level of belief in cooperation (where dark purple corresponds to high belief in cooperation and light purple corresponds to low belief in cooperation). The qualitative characteristics of each equilibrium are shown in the legend at the bottom of the figure (b). We characterize tightness, rate of cooperation and social homogeneity as ``high'' if they are above a threshold value of 0.95. (left) When norm utility weights are low ($\psi<0.5$), branching arises with high frequency in the Snowdrift game under descriptive norms, while intermediate levels of cooperation with high levels of conformity preference but without branching arises under descriptive norms. In the Prisoner's dilemma under descriptive norms, the defection and belief in defection evolve. (center) When norm utility weights are balanced ($\psi\approx0.5$), branching can arise in both the Snowdrift game and Prisoner's Dilemma under descriptive norms, but the population can also evolve towards either uniform defection or uniform cooperation. Under injunctive norms,  uniform defection and uniform cooperation can also arise. In addition in the Snowdrift game intermediate levels of cooperation with high levels of conformity preference but without branching arises, while in the Prisoner's Dilemma an equilibrium with high levels of conformity preference but low levels of cooperation arises. (right)) When norm utility weights are high ($\psi>0.5$) evolutionary branching is lost, while the remaining equilibria stay the same
c) The basin of attraction for equilibria under descriptive norms ($\theta=1)$ (top) and injunctive norms ($\theta=0$) bottom as a function of the norm utility weight $\psi$, for the Snowdrift game (left), and for the Prisoner's Dilemma. In cases where a type of equilibrium is not shown in a panel, its frequency is always zero. 
d) When norm utility weights are low in the Snowdrift game, two equilibria with low social homogeneity tend to emerge. This follows evolutionary branching, in which the population splits into two groups who differ in both their rate of cooperation, $p$, and their conformity preference, $\tau$. Plotted are the trait values for a population that undergoes branching, for each individual in an the population at each point in time.
Simulation results shown are ensemble averages from $10^4$ replicate simulations with uniformly distributed initial conditions (see Methods).  Simulations show a population of $N=100$ individuals, with benefit of cooperation $b=1/2$ and $c=1/3$ (Snowdrift game) or $c=3/4$ (Prisoner's Dilemma) and mutation rate $\mu=0.05$ and mutation effect size drawn uniformly from $\delta\mu\in[-0.05,0.05]$. Evolution occurs via an imitation process (see Methods) with selection intensity $s=10$. Simulations were run for $10^4$ generations, where each generation consists of $N$ imitation events.}
\end{figure}

Focusing on the Snowdrift game (i.e. $c<b$ -- we also show similar results hold for the Prisoner's Dilemma $b<c<2b$, see SI Section 2) we show that, when $\psi>0$, the system is bistable with two equilibria, which we label $+$ and $-$. The equilibrium $+$ corresponds to a population that uniformly expresses belief in cooperation ($q_+^*=1$) and cooperate with probability $p^*_+=\min\left[1,\frac{2((1-\psi)(b-c)+(1-\tau)\psi)}{(1-\psi)(2b-c)}\right]$. On the other hand, the state $-$ corresponds to a population that fully believe in defection $q^* = 0$ and cooperate with probability $p^*_-=\max\left[0,\frac{2((1-\psi)(b-c)-(1-\tau)\psi)}{(1-\psi)(2b-c)}\right]$. 

We can immediately make a number of observations about the nature of these equilibria. First, when only the material payoff contribute to utility $\psi=0$, $p^*_+=p^*_-=\frac{2(b-c)}{2b-c}$, which is the well-known Nash equilibrium for the Snowdrift game. Second, the equilibria depend on the preference for conformity over coherence, $\tau$, but not on the weight given to injunctive over descriptive norms, $\theta$. However, we do find that the stability of the equilibria depends on $\theta$ (see SI Section 2). Third, in general, we observe a misalignment between belief and behavior ($p_i \neq q_i$) for both equilibria.
\\
\\
\noindent \textbf{Classification of equilibria.}
 We identify the degree of alignment between belief and behavior with cultural tightness, which has been widely studied in the empirical literature \cite{gelfand2011differences,harrington2014tightness}. Intuitively, a population has a high degree of cultural tightness if its members share a common belief (i.e., the norm is \textit{shared}) and behave in accordance with that belief (i.e., the norm is \textit{strong}). We therefore measure the cultural tightness of a population as

\begin{eqnarray}
  \nonumber  \mathcal{T}=
    &\left( \sum_{i \neq j} \frac{q_iq_j+(1-q_i)(1-q_j)}{N(N-1)} \right)\times\\
    &\left( \sum_{i \neq j} \frac{p_iq_j+(1-p_i)(1- q_j)}{N(N-1)} \right)
\label{eq tightness2}
\end{eqnarray}

The first term describes the probability that two individuals express the same belief. The second term describes the probability that one individual acts in accordance with the expressed belief of another. Accordingly, $\mathcal{T} \in [0,1]$, and tightness is minimized, $\mathcal{T}=0$, only if $p_i = 0$ and $q_i = 1$ (or vice-versa) for all members of the population, $i$. Tightness is maximized, $\mathcal{T}=1$, if $q_i = p_i=0$ or $q_i=p_i = 1$ for all members of the population, $i$.

We can write 
$\mathcal{T} = \mathcal{T}_b \mathcal{T}_a$
where $\mathcal{T}_b$ is the first term in eq~\ref{eq tightness2}, measuring the tightness of belief (i.e. the extent to which members of the population believe the same thing) and $\mathcal{T}_a$ measures tightness of action and beliefs (i.e. the extent to which individuals' actions accord with the beliefs of others).
We can further define coherence with self (i.e. the extent to which an individuals' actions accord with their own  beliefs)
\begin{equation*}
    \mathcal{C}= \left( \sum_{i } \frac{q_i p_i+(1-p_i) (1-q_i)}{N} \right)
\label{eq cohese}
\end{equation*}

We can now classify the tightness of the population as follows: 

\begin{itemize}
\item Tight populations. $\mathcal{T}_b=\mathcal{T}_a=\mathcal{C}=1$ (Figure 2 and 3, filled dots)
\item Loose populations with commonly shared but incoherent beliefs. $\mathcal{T}_b=1$, $\mathcal{T}_a<1$, $\mathcal{C}<1$ (Figure 2 and 3, empty dots and Figure 2 heterogeneous populations)
\item Loose populations with heterogeneous but coherent beliefs. $\mathcal{T}_b<1$, $\mathcal{T}_a<1$, $\mathcal{C} \leq 1$ (Figure 3, heterogeneous populations).
\end{itemize}
Note that a state with $\mathcal{T}_b<1$ and $\mathcal{T}_a=1$ is not possible, since actions cannot always be consistent with other's beliefs in a population where beliefs are heterogeneous. 


We observe the first three of these states in populations where belief, behavior and preferences are allowed to co-evolve (see Figure 2-3 below). When preferences are fixed, and only belief and behavior coevolve, matters are more simple. We can easily calculate the degree of cultural tightness at the two (monomorphic) equilibria identified above. Here $\mathcal{T}_+=p^*_+=\min\left[1,\frac{2((1-\psi)(b-c)+(1-\tau)\psi)}{(1-\psi)(2b-c)}\right]$ and $\mathcal{T}_-=1-p^*_-=\min\left[1,\frac{2((1-\psi)c/2+(1-\tau)\psi)}{(1-\psi)(2b-c)}\right]$. Thus, if $c<2b/3$ then $\mathcal{T}_+>\mathcal{T}_-$ and a population that evolves to believe in cooperation is tighter than a population that evolves to believe in defection.


In this example, the population is monomorphic at equilibrium -- meaning all members of the population evolve to have similar rates of cooperation, and similar levels of belief in cooperation. Nonetheless, low levels of cultural tightness can still arise due to the mismatch between individual belief and behavior. An alternate scenario occurs in heterogeneous populations, in which the population is non-monomorphic at equilibrium, and instead splits into distinct groups, which in general may have different beliefs, behaviors and preferences, and therefore lead to low overall cultural tightness at the level of the population. 

To account for these different scenarios, in addition to the degree of cultural tightness (Eq. 3), we characterize populations in terms of their \emph{social homogeneity}, i.e. the extent to which the population is split into distinct groups.
We restrict ourselves to scenarios in which the population splits into two groups such that the degree of social homogeneity can be measured simply as

\begin{equation}
\mathcal{S}=1-\frac{n^{\text{out}}}{n^{\text{in}}}
\label{eq social homo}
\end{equation}

where $n^{\text{in}}$ is the number of individuals belonging to the dominant (i.e. largest) social group and $n^{\text{out}}$ is the number of individuals belonging to the smaller group. When there is only a single group, as in the example above, $n^{\text{out}}=0$, and social homogeneity is maximized ($\mathcal{S}_{1}=1$). We determine whether the population has split into two groups based on whether evolutionary branching has occurred, such that distinct groups who cooperate at different rates have been maintained for at least $10^3 N$ update steps in simulations (see SI Section 3 for further details).

Putting this all together, we classify the equilibrium state of a population in terms of two emergent societal properties, in addition to the level of belief in cooperation, and the rate of cooperation itself: i) the degree of cultural tightness of the population and ii) the degree of social homogeneity in the population. When preferences are allowed to evolve alongside belief and behavior there are often multiple stable equilibria, which differ in multiple characteristics. To simplify matters we classify the level of a characteristic as either ``high'' or ``low'' (see Figure 2 and 3) when contrasting them with one another. We also provide a detailed breakdown of the quantitative characteristics for each equilibrium in the SI Section 3 (Figure S4-S8).
\\
\\
\noindent \textbf{Evolution of conformity preference.}
We can now explore the evolution of preferences alongside behaviors and beliefs. We begin by considering the evolution of $\tau$ -- an individual's preference for conformity over coherence -- alongside the evolution of cooperation ($p$) and belief in cooperation ($q$), while keeping the other parameters ($\psi$ and $\theta$) fixed.

Invasion analysis (see SI Section 2) reveals a proliferation of equilibria (see Figure 2) when conformity preference is allowed to evolve, so that in general the system permits six qualitatively different equilibria, whose basin of attraction and stability depend critically on the overall norm-utility weight, $\psi$, and the extent to which conformity occurs via descriptive or injunctive norms, $\theta$. We explore this rich phenomenology by systematically varying the norm utility weight $\psi$, under either completely descriptive norms ($\theta=1$) or completely injunctive norms ($\theta=0$). 

A first pair of evolutionary equilibria corresponds to states of an highly-tight coherent population, where all individuals believe in the same thing and act accordingly to it ($q^*=p^*$), with strong preference for coherence over conformity (small $\tau$). These two states correspond to the case of full cooperation ($q^*=p^*=1$) or full defection ($q^*=p^*=0$). For both games considered here (Prisoner's dilemma and snowdrift) and for both injunctive and descriptive norms, these two states emerge for large enough norm utility weight (large enough $\psi$).

Another class of evolutionary states corresponds to strategies ($p^*$) matching the Nash equilibrium ($p^*=0$ for Prisoner dilemma, and $p^*=1/2$ for Snowdrift game) and strong conformity preference ($\tau=1$). In this case, the population has homogeneous beliefs (in cooperation only $q^*=1$ for Prisoner dilemma and in either cooperation $q^*=1$ or defection $q^*=0$ for snowdrift game), but incoherent behavior. These states emerge only for injunctive norms ($\theta=0$), where conformity is only determined by beliefs, and for any norm utility weight.
In these states, the population acts as Nash, while maximizing conformity of beliefs, effectively decoupling the contribution of strategies (which only affect material payoff, therefore converging to the Nash equilibria) from the beliefs (which, in absence of coherence, are the only contributor to norm utility).

A third class of equilibrium states correspond to heterogeneous populations (branched equilibria). In this case, a population partitions into two groups of individuals: one class  has a high preference for coherence ($\tau=0$) and acts in accordance of their belief ($q_{\tau=0}^*=p_{\tau=0}^*$), the other, lower abundant, class has high preference for conformity ($\tau=1$), have the same belief of the other class while acting in the opposite way ($1-p_{\tau=1}^*=q_{\tau=1}^*=p_{\tau=0}^*$). This class of equilibrium emerges only for descriptive norms ($\theta=1$) where the incentive is to adapt the belief to the action of others and for small enough norm utility. In the snowdrift game both $p^*_{\tau=0}=0$ and $p^*_{\tau=0}=1$ are possible, while in the Prisoner dilemma only $p^*_{\tau=0}=1$ is possible.

The branched equilibria not only exhibit a loss of overall cultural tightness compared to the monomorphic equilibria, but also a loss of social homogeneity, with distinct groups emerging who have different preferences and behaviors. The conditions for evolutionary branching to occur are described in detail in the SI (section 2). We show that belief $q$ is always bi-stable for any value of $\theta$ (where intermediate values of $\theta$ correspond to a mixture of injunctive and descriptive norms), such that belief always evolves to its maximum ($q=1$) or minimum ($q=0$). Branching can then occur in the remaining two traits, i.e. the probability of cooperation, $p$, and the conformity preference $\tau$, as described above. We show that selection on $p$ and $\tau$ will be diversifying and lead to branching provided norms are purely descriptive ($\theta=1$), and provided the initial tightness of the population is not too great (i.e. cooperation and belief in cooperation are sufficiently diverged from one another).

\begin{figure}[tbhp]
\centering
\includegraphics[width=0.8\linewidth]{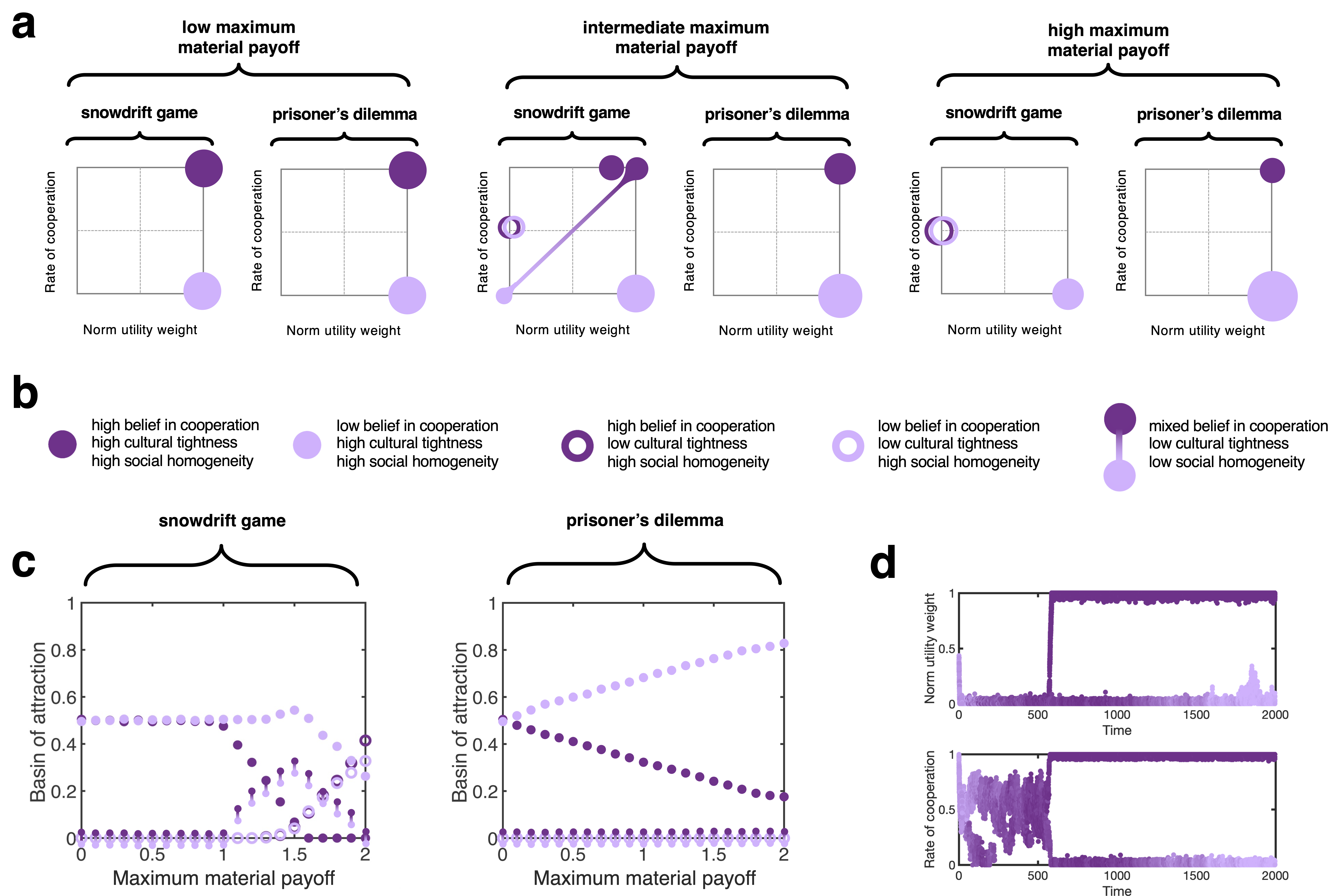}
\caption{\footnotesize \textbf{Co-evolution of cooperation, belief and norm utility weight.} We performed individual-based simulations for the co-evolution of cooperation, $p$, belief in cooperation, $q$ and the norm utility weight $\psi$, for high levels of coherence preference ($\tau=0$). The procedure is the same as described for Figure 2. a) Qualitative characteristics of the different types of equilibria. (left) When maximum material payoff is low ($b<1$), the population is bistable, with high norm utility weight, high cultural tightness and either high or low belief in cooperation . (center) When the maximum material payoff takes an intermediate value ($1<b<2$), branching can arise in the Snowdrift game, but the population can also evolve towards either uniform defection or uniform cooperation. (right) When maximum material payoff is high high ($b\gtrapprox 2$) evolutionary branching is lost in the Snowdrift game, and the norm utility weight tends to evolve to lower values, with corresponding loss of cultural tightness. Note that the equilibria for the Prisoner's Dilemma are qualitatively unchanged throughout. c) The basin of attraction for equilibria under low conformity preference, ($\tau=0)$  as a function of the maximum material payoff $b$ (keeping the ratio $c/b$ fixed), for the Snowdrift game (left), and for the Prisoner's Dilemma. d) When the maximum material payoff is sufficiently large in the Snowdrift game a new equilibrium with low social homogeneity tends to emerge. This follows evolutionary branching, in which the population splits into two groups who differ in both their rate of cooperation, $p$, and their norm utility weight, $\psi$. This is followed by divergence in the level of belief in cooperation between the two groups. Plotted are the trait values for a population that undergoes branching, for each individual in an the population at each point in time. Note that  here level of belief is indicated by the color of the data point (dark purple for high belief in cooperation, light purple for low belief) and changes over time.  
Simulation results shown are ensemble averages from $10^4$ replicate simulations with uniformly distributed initial conditions (see Methods).  Simulations show a population of $N=100$ individuals, with benefit of cooperation varying and the cost of cooperation set to $c=b/3$ (Snowdrift game) or $c=3b/4$ (Prisoner's Dilemma) and mutation rate $\mu=0.05$ and mutation effect size drawn uniformly from $\delta\mu\in[-0.05,0.05]$. Evolution occurs via an imitation process (see Methods) with selection intensity $s=10$. Simulations were run for $10^4$ generations, where each generation consists of $N$ imitation events.}
\end{figure}

  While our analysis shows (see SI section 2) that branching of this type occurs when norms are purely descriptive ($\theta=1$), we also find by individual-based simulation that this type of branching can occur when norms are mixed ($0<\theta<1$) and, in particular, we find that as the use of descriptive norms becomes more frequent, branching becomes more likely (Figure S6).
\\
\\
\noindent \textbf{Evolution of norm-utility weight.}
Next we consider the evolution of $\psi$ -- the weight given to norm utility -- alongside the evolution of cooperation ($p$) and belief in cooperation ($q$), while keeping the other parameters ($\tau$ and $\theta$) fixed. Invasion analysis (see SI Section 2) reveals that in general the system permits five qualitatively different stable equilibria, whose basin of attraction depends critically on the material payoff of the game, $b$ (i.e the benefit of cooperation). To illustrate this, we systematically vary the benefit of cooperation $b$, while keeping the ratio $b/c$ fixed, such that the game associated with the material payoffs remains fixed. We then evaluate the case of complete conformity preference ($\tau=1$) and complete coherence preference ($\tau=0$).

One class of equilibria corresponds to fully coherent individuals ($p^*=q^*$) with high norm utility weight ($\psi^*=1$). These two states (corresponding to cooperation or defection) emerge for a wide range of the maximum material payoff (determined by $b$) and for both games.

Another class of equilibria corresponds to low norm utility weight ($\psi^*=0$),
meaning that players only take account of material payoff in determining their utility, and the equilibria of the system are the same as those for the material game without belief. These equilibria appear when the maximum material payoff is high, and for both injunctive norms ($\theta=0$) descriptive norms ($\theta=1$), and for both high conformity $(\tau=1)$ and high coherence $(\tau=0)$ preferences (see Figure 3)



When the maximum material payoff is high and the preference favor coherence ($\tau=0$), we see a new branched equilibrium emerge, along with lower norm utility weights and lower levels of cultural tightness (Figure 3).
This branched equilibrium is qualitatively different from those that occur when only $\tau$ is allowed to evolve. The conditions for branching to occur are described in detail in the SI (section 2). We show that initially the system evolves to a monomorphic state such that belief corresponds to defection ($q=0$). Branching can then occur in the remaining two traits, i.e. the probability of cooperation, $p$, and the norm-utility weight $\psi$. We show that selection on $p$ and $\psi$ can be diversifying and lead to branching provided the material payoffs satisfy $b>\frac{c/b}{2(1-c/b)}$, i.e. provided the maximum material payoff is large enough, given a fixed game (fixed $c/b$).
When branching occurs the population splits into two groups with one group exhibiting cooperative behavior (large $p^*$) alongside a high norm utility  weight (large $\psi$) and a high level of belief in cooperation (large $q$), and the other group exhibiting no cooperative behavior (low $p$), alongside a low norm utility weight (low $\psi$) and a low level of belief in cooperation (in particular, belief in cooperation is released from selection in the second group, see SI section 2).

In sum, depending on which preferences are able to evolve, quite different types of groups can arise through evolutionary branching. When conformity preference $\tau$ is allowed to evolve, the population is sensitive to whether norms are descriptive or injunctive, tends to maintains a common belief, but can split into groups that behave differently, and that have different preferences for conformity vs coherence of belief. When norm-utility weight is allowed to evolve, the population splits into groups who differ in both their behavior, their beliefs and their preferences.
\\
\\
\noindent \textbf{Response to shifting environments.}
We now explore how the emergent societal properties of a population change in response to a shifting material environment, when all three preferences -- conformity preference ($\tau$), norm utility weight ($\psi$), and injunctive norm weight ($\theta$) -- are allowed to co-evolve alongside belief and behavior.

We explore two kinds of shift in the material environment. We consider a shift in the relative costs and benefits of cooperation, $c/b$, while keeping the maximum material payoff, $b$, fixed. We also consider a shift in the maximum material payoff $b$, while keeping the ratio $c/b$ fixed.

We characterize the change in social homogeneity, cultural tightness and rate of cooperation in the population in response to a changing material environment (Figure 4). We see that a more challenging material environment (i.e. lower values of $b$ or higher values of $c/b$) is associated with lower cultural tightness, and lower cooperation, consistent with previous results \cite{gelfand2011differences}. However we also see that changes in cultural tightness are often associated with fragmentation through loss of social homogeneity, in which the population undergoes branching into two distinct groups, especially for intermediate benefits of cooperation (Figure 3).

\begin{figure}[tbhp]
\centering
\includegraphics[width=0.9\linewidth]{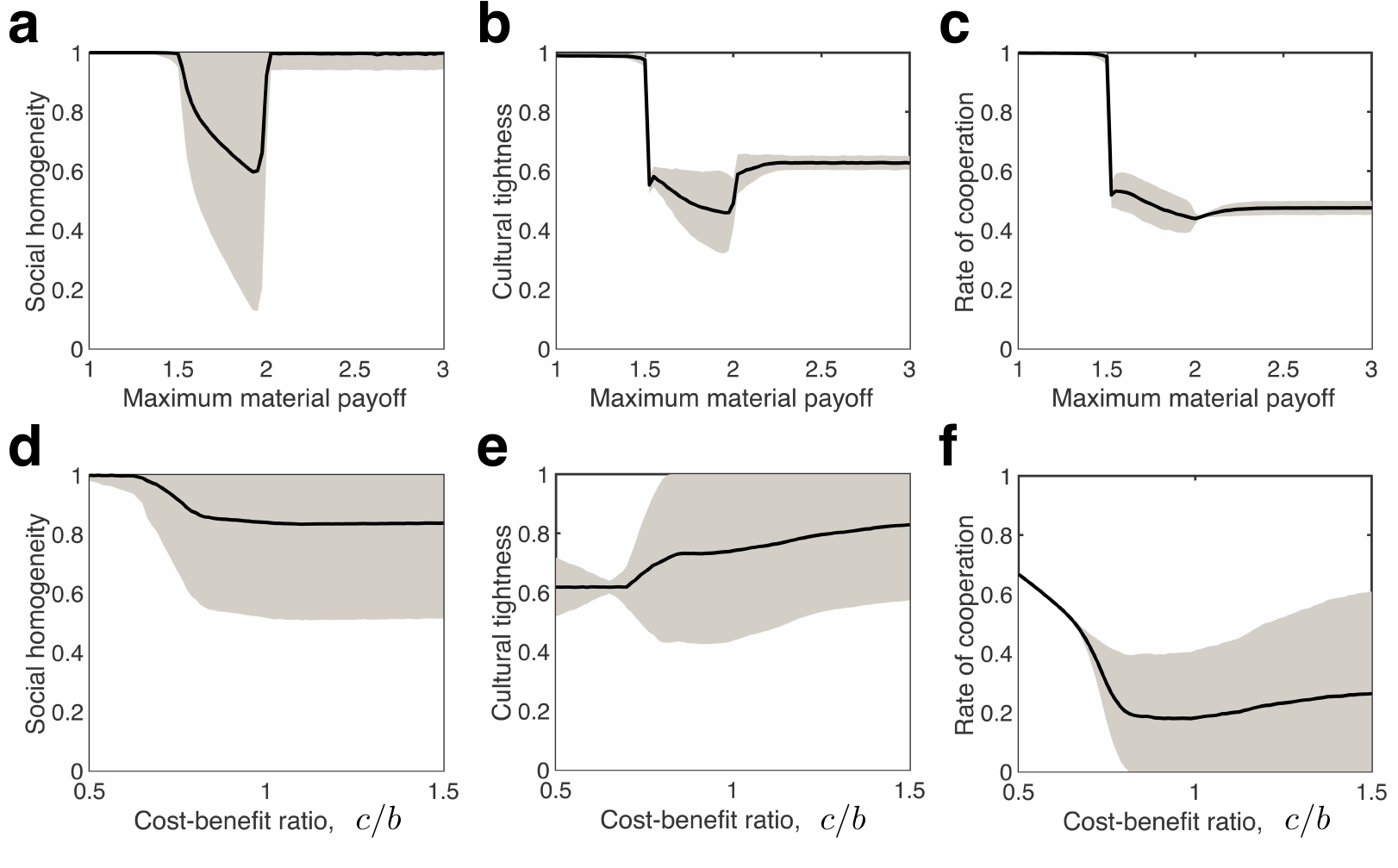}
\caption{\footnotesize \textbf{Shifting material environment}. We studied the effect of changing the maximum material payoff, $b$ while keeping the game fixed such that $c=2b/3$ remains constant and the game is always a Snowdrift game (a-c). The population is initialized close to a cooperative equilibrium such that $p=q=1$  for each member of the population, along with $b=1$. Evolution then occurs for $10^4N$ updates, before the environment starts to shift. We then increase the maximum material payoff in increments of 0.025 every $1000N$ updates until $b=3$. We allow evolution to occur in all five traits in the model, i.e. the rate of cooperation $p$, belief in cooperation $q$, conformity preference $\tau$, norm utility weight $\psi$ and descriptive norm weight $\theta$. We characterize the state of the population in terms of (a) social homogeneity, (b) cultural tightness and (c) rate of cooperation. We see that an improving material environment (increasing $b$) leads to a decrease in cultural tightness and, paradoxically, a decrease in the rate of cooperation. For intermediate levels of cooperation benefit, there is a sharp decline in social homogeneity, corresponding to frequent evolutionary branching (see Table 1 for details of the branched groups). 
We then studied the effect of changing the ratio $c/b$ while keeping the maximum material payoff fixed $b=3/2$ (d-f). We see that a declining material environment (decreasing $c/b$) leads to a decrease in social homogeneity (c), an increase in cultural tightness (d) and a decrease in the rate of cooperation. Note that the game is a Prisoner's Dilemma when $c/b>1$ and a Snowdrift game otherwise. 
Simulations show a population of $N=100$ individuals, with mutation rate $\mu=0.05$ and mutation effect size drawn uniformly from $\delta\mu\in[-0.05,0.05]$. Evolution occurs via an imitation process (see Methods) with selection intensity $s=10$.}
\end{figure}

When branching occurs, it is associated with differences between groups in rates of cooperation, belief in cooperation, norm-utility weight, conformity preference and injunctive norm weight (Table 1).

\begin{table}
\centering
\caption{Group characteristics following branching, for a Snowdrift game with $b=3/2$ and $c=1$ in a population of $N=100$.}
\begin{tabular}{lrrr}
Average trait & Group 1 & Group 2 \\
\hline
Cooperation rate ($p$) & $0.999\pm 0.005$ & $0.003 \pm 0.009$  \\
Belief in cooperation ($q$) & $0.999\pm 0.005$ & $0.56\pm 0.259$  \\
Conformity preference ($\tau$) & $0.003 \pm 0.009$ & $0.758\pm 0.134$  \\
Norm utility weight ($\psi$) & $0.996\pm 0.01$ & $0.002\pm 0.007$  \\
Injunctive norm weight ($\theta$) & $0.309\pm 0.178$ & $0.619\pm 0.189$  \\
Group size & $0.645\pm 0.007$ & $0.335\pm 0.007$  \\
\hline
\end{tabular}

\end{table}

In particular we see one group, which has a high norm-utility weight, engages in cooperation, believes in cooperation and has a preference for belief coherence rather than conformity, evolves alongside a second group that has low norm utility weight and does not cooperate. Since the second group does not derive utility from conformity or coherence, the trait values for belief, conformity preference and norm utility are released from selection (see SI Figure S7-S8). These groups can broadly be characterized as behaving in a normative way (group 1), with cooperation taking place without regard for material payoff, or else behaving in a ``rational'' way (group 2), with little cooperation taking place, and little regard for norm utility. 

The emergence of such groups, and the associated loss of social coherence, tends to occur when the maximum material payoff takes intermediate values, and can persist for both the Snowdrift game ($2c<b<c$) and the Prisoner's Dilemma ($b>c$).

\section*{Discussion}

The feedback between beliefs, behaviors, and preferences shapes individuals' response to opportunities for cooperation in a way that is not adequately captured by the study of just beliefs or behaviors alone. Individual attitudes in turn shape the ability of societies to adapt and respond to threats, by creating a culture that is tighter or more fragmented around questions of cooperation and collective action.  
Here we develop a framework for studying the coevolution of beliefs, behaviours, and preferences in a game-theoretic setting. We assume that individuals’ utilities combine a material and a psychological component (norm utility), with the material component derived from the costs and benefits of cooperation. 
We analyze the evolutionary dynamics of beliefs, behaviors and preferences, and show that, depending on which preferences are able to evolve, this leads to a proliferation of the number of stable equilibria compared to the single equilibrium of the corresponding classical game. 
Finally, we show that when all preferences are allowed to evolve, shifts in the material environment (i.e. changes in the costs and benefits of cooperation) can drive the emergence of both tighter and more fragmented cultures. 
Our modeling approach makes use of three novel elements: (i) the incorporation of beliefs and norm utility into an expanded normal form payoff matrix, (ii) the endogenous evolution of norm-utlity preferences, and (iii) the formal definition and analysis of the evolution of cultural tightness and social homogeneity.

Classical game theory assumes that players are solely motivated by material rewards, which depend on the actions of other players. More refined game-theoretic approaches deviate from this assumption of strict rationality, intended as optimal inference and material payoff maximization, and instead account for imperfect information, cognitive biases and beliefs in decision-making processes \cite{bicchieri2004rationality, dekel2015epistemic}. For example, some models perturb the normal form matrix of classical games by adding an additional reward for norm compliance \cite{bicchieri2011social}. In such \textit{mixed-motive} games, where the unique Nash equilibrium results in a suboptimal outcome (as in the Prisoner's dilemma) the existence of a norm can transform the payoff matrix into that of a simple coordination game: this would resolve the dilemma, facilitate individuals' choices and favor collective interest \cite{bicchieri2005grammar}. 
However, treating the psychological reward of norm compliance as fixed fails to account for the dynamic nature of beliefs \cite{castellano2009statistical}. 
Recent norm-utility models \cite{calabuig_culture_2018, gavrilets_coevolution_2021} consider the evolution of belief dynamics using DeGroot-like updating processes: beliefs do not explicitly enter into the payoff matrix, which remains unchanged from the classical game. In contrast, our model explicitly considers belief expression as an action that couples with the standard ``material'' one. This coupling expands the dimensionality of the normal form matrix. In our model, expressed belief refers to the individual’s support for the norm (in our case, cooperation). This expressed belief may be influenced by, but does not coincide with, \textit{empirical expectations} (beliefs about norm adherence among the interacting agents) or \textit{normative expectations} (beliefs about others' injunctions) \cite{bicchieri2023s, barrett2021some}. Our approach has both advantages and limitations. On the one hand, our model focuses on two variables that are potentially observable and measurable, -- expressed belief and material action -- rather than more internal and ambiguous states such as subjective expectations. On the other hand, by assuming that belief expression depends solely on the behaviors and/or expressed beliefs of others, as well as on one’s own behavior, we neglect subjective and potentially strategic perceptions, as well as the influence of individuals’ personal values \cite{elster2021guiding, dalege2024networks}, all of which likely play a role in belief expression.

Whether people focus more on observed behaviors or on what others believe should be done, so on descriptive or injunctive norms, has a key influence on norm adoption, spread and change \cite{bicchieri2014norms}. It has been argued that inferring what others consider appropriate (injunctive norms) requires more cognitive effort than merely observing their behavior \cite{kredentser2012following}. For this reason, Morris et al. \cite{morris2015normology} emphasize that descriptive norms often lead to automatic, socially safe behavioral responses, whereas attention to injunctive norms functions more like a ``social radar'', requiring continuous monitoring of others’ attitudes. However, behaviors may be observable only in limited contexts (for example, private, domestic actions), and injunctive norms can guide behavior across a broader range of situations. Thus, individuals' reliance on descriptive versus injunctive norms may be driven by informational availability and personal willingness to engage in cognitive effort. In this sense, providing additional information can shift individuals’ reliance and potentially catalyze norm change, though such interventions of norm nudging may also have unintended effects \cite{bicchieri2022nudging}. Such unintended consequences can be seen in our model, where a shift from injunctive to descriptive norms can, in some contexts, spark social fragmentation.  Further complexity arises from voluntary distortions in belief expression \cite{bicchieri2023s}, or misperception of others’ beliefs (as in the case of misperception of public support for climate action \cite{sparkman2022americans}), both of which have been shown to significantly shape belief dynamics \cite{galesic2021integrating, zimmaro2025meta}, and are not well captured by our model.

By considering the evolution of the preference parameter $\tau$ regulating the weight given to conformity over coherence, $\psi$ regulating the weight given to psychological vs material payoff, and $\theta$ regulating the weight given to descriptive vs injunctive norms, we account for the fact that individuals may learn from their neighbours not only \textit{what to think} (i.e., adopt their belief), but also \textit{how to think}, i.e., how much relevance and attention to assign to different factors influencing their decision-making. Considering these adaptive weights is in line with other well-studied adaptive processes in the literature, such as adaptive networks, where individuals sever (or disregard) social ties with disagreeing peers \cite{holme2006nonequilibrium, iniguez2009opinion}. Another example is the disagreement-reinforcement mechanism characteristic of echo chambers: individuals trapped in an echo chamber not only adopt the radical views of their group, but also internalize a psychological mechanism by which antagonistic beliefs are systematically discredited, such that challenging them may even strengthen existing convictions \cite{nguyen2020echo}.
In this sense, the tendency to be coherent vs. conformist, or materialistic vs. ``moralistic'' can be conceptualized in terms of evolving meta-norms with respect to a focal norm of cooperation \cite{gelfand2024norm}. Indeed, the relative importance of showing coherence as opposed to conformity, for example, is something that can be socially learned, just like other auxiliary meta-norms (for instance, those regulating whether or not to punish norm violators, whose evolution has been considered in previous models of norm dynamics \cite{axelrod1986evolutionary, gavrilets2012evolutionary}).

The analysis of our model shows that the addition of beliefs to the classical payoff matrix and the evolution of the internal parameters influencing the decision-making process leads to a proliferation of equilibria compared to classical games. Moreover, we find that under certain conditions the system undergoes evolutionary branching, whereby an initially homogeneous population fragments through social learning into two or more groups with different beliefs, behaviors, and preferences. Naively one might expect loss of social homogeneity to be accompanied by a reduction in cultural tightness, as groups emerge with different beliefs, behaviors and preferences. However we find a more complex interplay, in which loss of social homogeneity can either be accompanied by increase in cultural tightness on average, or the converse. The precise nature of the relationship depends on the nature of the material game being played \cite{doi:10.1073/pnas.2102140118,doi:10.1126/sciadv.abd4201}. 

In summary, our work highlights several key factors that can influence the evolution of norm adherence, which in turn contributes to the overall level of cultural tightness and social cohesion in a society. These factors are (a) the structure of the material game being played, (b) the type of information used when following norms (i.e. are norms injunctive, descriptive or mixed), and (c) the extent to which individual preferences for coherence versus conformity and material versus psychological reward are able to evolve through social learning. Understanding the connections between these individual-level social-psychological mechanisms and macroscopic properties that emerge at the collective level \cite{holme2015mechanistic, sontuoso2024mathematical}, such as cultural tightness and social cohesion, can help us to better relate belief dynamics, the evolution of social norms, and collective action, thereby improving our understanding of how societies evolve, adapt and respond to threats.

\section*{Methods}
 Here we describe the method of transforming normal form payoff matrices to incorporate beliefs, and a two-trait model for the co-evolution of belief and behavior.
\\
\\
\noindent \textbf{Payoff matrix transformation.}
For the standard $2\times 2$ payoff matrix in a game with actions \emph{cooperate} ($CO$) and \emph{defect} ($DE$) we have a utility contribution (which we refer to as the \emph{material payoff}) of the form 
\begin{equation}
    \pi^{\text{mat}}  : \{CO, DE\}  \times \{CO, DE\} \to \mathbb{R}
\end{equation}
\begin{eqnarray}
  \nonumber  \pi^{\text{mat}}(X,Y) = 
 \nonumber   \begin{cases} 
    a_{CO,CO} \quad \quad X=CO,Y=CO\\
    a_{CO,DE} \quad \quad X=CO,Y=DE\\
    a_{DE,CO} \quad \quad X=DE,Y=CO\\
    a_{DE,DE} \quad \quad X=CO,Y=DE
    \end{cases}\\
\end{eqnarray}

and $a_{X,Y}$ is the payoff received by the focal player, given his action $X$ and the action of his opponent $Y$ (i.e. these are the payoffs of the standard normal form game). In the Snowdrift setup considered in the main text we have $a_{CO,CO}=b-c/2$, $a_{CO,DE}=b-c$, $a_{DE,CO}=b$ and $a_{DE,DE}=0$, where $b$ and $c$ are the costs and benefits of cooperation.

This material payoff is supplemented by a \emph{psychological payoff}, or norm utility, derived from from both the actions of the individuals and their expressed beliefs. In our model, expressed beliefs are simply binary and take the form of \emph{belief in cooperation} ($co$) and \emph{belief in defection} ($de$). 

The resulting utility contribution of the psychological payoff is
\begin{equation}
    \pi^{\text{psych}}  : \{CO,DE\}\times\{CO,DE\}\times\{co, de\}  \times \{co, de\} \to \mathbb{R}
\end{equation}
\begin{eqnarray}
 \nonumber   &\pi_{psych}(X,Y,x,y) = \\
   \nonumber & \begin{cases} 
    \beta^{co,co}_{CO,CO} \quad  X=CO,Y=CO,x=co,y=co \\
    \beta^{co,de}_{CO,CO} \quad  X=CO,Y=CO,x=co,y=de \\
    \beta^{de,co}_{CO,CO} \quad  X=CO,Y=CO,x=de,y=co \\
    \beta^{de,de}_{CO,CO} \quad  X=CO,Y=CO,x=de,y=de \\
    \beta^{co,co}_{CO,DE} \quad  X=CO,Y=DE,x=co,y=co \\
    \beta^{co,de}_{CO,DE} \quad  X=CO,Y=DE,x=co,y=de \\
    \beta^{de,co}_{CO,DE} \quad  X=CO,Y=DE,x=de,y=co \\
    \beta^{de,de}_{CO,DE} \quad  X=CO,Y=DE,x=de,y=de \\
    \beta^{co,co}_{DE,CO} \quad  X=DE,Y=CO,x=co,y=co \\
    \beta^{co,de}_{DE,CO} \quad  X=DE,Y=CO,x=co,y=de \\
    \beta^{de,co}_{DE,CO} \quad  X=DE,Y=CO,x=de,y=co \\
    \beta^{de,de}_{DE,CO} \quad  X=DE,Y=CO,x=de,y=de \\
    \beta^{co,co}_{DE,DE} \quad  X=DE,Y=DE,x=co,y=co \\
    \beta^{co,de}_{DE,DE} \quad  X=DE,Y=DE,x=co,y=de \\
    \beta^{de,co}_{DE,DE} \quad  X=DE,Y=DE,x=de,y=co \\
    \beta^{de,de}_{DE,DE} \quad  X=DE,Y=DE,x=de,y=de \\
    \end{cases}\\
\end{eqnarray}
where $\beta^{x_i,x_j}_{X_i,X_j}$ stands for the psychological utility of the focal player, given his action $X$, his expressed belief $x$, and the interacting individual's action $Y$ and expressed belief $y$. In general we may choose the 16 $\beta$ parameters in any way we like, to capture the effects of different types of social norm and the resulting norm utility. As described in the main text, we focus on three different sources of norm utility: 1) utility derived from coherence between beliefs and actions, 2) utility derived from conforming one's belief with the belief of others (injunctive norms) and 3) utility derived from conforming one's belief with the actions of others (descriptive norms). This results in the following parameterization (see Eq. 2)

$$
\beta^{x_i,x_j}_{X_i,X_j}=(1-\tau_i)\delta_{x_iX_i}+\tau_i\left(\theta_i\delta_{x_iX_j}+(1-\theta_i)\delta_{x_ix_j}\right)
$$

where $\delta_{x_iX_i}$ is a Kronecker delta such that 

\begin{eqnarray}
  \nonumber  \delta_{x_iX_i} = 
 \nonumber   \begin{cases} 
    1 \quad \quad x_i=co,X_i=CO\\
    0 \quad \quad x_i=co,X_i=DE\\
    0 \quad \quad x_i=de,X_i=CO\\
    1 \quad \quad x_i=de,X_i=DE
    \end{cases}  
\end{eqnarray}

i.e. it is equal to 1 when belief $x_i$ and behavior $X_i$ coincide, and 0 otherwise. Similarly $\delta_{x_ix_j}$ is 1 when beliefs $x_i$ and $x_j$ coincide, and so on.

Combining the material and psychological utility according to Eq 1. then produces the $4\times4$ payoff matrix shown in Figure 1.
Going beyond binary beliefs and actions, if one considers $S_A$ and $S_b$ respectively the set of all the possible actions and beliefs, with cardinality respectively $n_A = |S_A|$ and $n_b = |S_b|$, the utility function with pairwise interactions will be of the type
\begin{equation}
    \pi  : S_A \times S_b \times S_A \times S_b \to \mathbb{R},
\end{equation}
and induce a $n_An_b \times n_An_b$ payoff matrix.
\\
\\
\noindent \textbf{Multi-trait model.} Having transformed the $2\times2$ payoff matrix for the Snowdrift game/Prisoner's Dilemma to account for belief (Figure 1), we analyze the evolutionary dynamics of behavior and belief (two-trait model) as well as the dynamics when preferences for coherence and conformity are also allowed to evolve. To do this we first construct a ``two-trait'' model, in which, for each pairwise interaction they engage in, an individual $i$ independently chooses the material action ($\{CO,DE\}$) with probability $p_i$ and the belief they express ($\{co,de\}$) with probability $q_i$. In addition, the individual has an initial relative preference for conformity over coherence (meta-norm) $\tau_i$. The other parameters, $b$, $c$, $\psi$ and $\theta$ are assumed to be determined exogenously, and are the same for all individuals. 

Under this model, the probability of the individual's composite action $(CO,co)$ is thus $p_iq_i$, the probability of $(CO,de)$ is $(1-p_i)q_i$, the probability of $(DE,co)$ is $p_i(1-q_i)$ and the probability of $(DE,de)$ is $(1-p_i)(1-q_i)$. We assume that an individual's overall utility is derived from interactions with each member of a large population of size $N$, i.e.

$$
\pi_{i}(p_i,q_i)=\frac{1}{N-1}\sum_{j\neq i} \pi_{ij}(p_i,q_i,\tau_i,p_j,q_j)
$$

The payoff for a given interaction between two players is given by Eqs. 1-2 and the utility is written out in full detail and analyzed in the SI Section 1. We also analyze a ``three-trait'' model in which the preference for conformity vs cohesion are allowed to evolve alongside $p$ and $q$.

In order to determine the equilibria of the evolutionary dynamics we perform evolutionary invasion analysis [REF], in which we assume that populations are very large, $N\to\infty$, while mutations -- which change the value of evolving traits -- have small effects. Taking the example of the three-trait model, we  calculate the selection gradient associated with a mutant strategy $(p_i,q_i,\tau_i)$ in a population where all other individuals use a resident strategy $(p_r,q_r,\tau_r)$, that is we calculate $\frac{\partial \pi_{i}}{\partial p_i}\big |_{p_i=p_r}$, $\frac{\partial \pi_{i}}{\partial q_i}\big |_{q_i=q_r}$ and $\frac{\partial \pi_{i}}{\partial \tau_i}\big |_{\tau_i=\tau_r}$. If the selection gradient is zero, the resident strategy $(p_r,q_r,\pi_r)$ is an equilibrium of the evolutionary dynamics. We calculate the location of all equilibria, and assess their stability, in SI Section 2.
\\
\\
\noindent \textbf{Definition of tightness.} 
Taking the definition given in \eqref{eq tightness2}, we can rewrite the two terms $\mathcal{T}_b, \mathcal{T}_a$ as
\begin{align*}
    & \mathcal{T}_b =  1-2[\langle q \rangle (1-\langle q \rangle) ] - \frac{2}{N-1}Var(q)\\
    & \mathcal{T}_a =  1-[\langle q \rangle (1-\langle p \rangle) + \langle p \rangle (1-\langle q \rangle)] - \frac{2}{N-1}Cov(q,p)
\end{align*}
where $\langle q \rangle $, $\langle p \rangle $ are respectively the average belief and behaviour. Analyzing the belief-related term $\mathcal{T}_b$, we see that it depends primarly on the extent to which beliefs are mixed, $\langle q \rangle (1-\langle q \rangle)$, while the contribution of the dispersion of beliefs $Var(q)$ is subleading, of order $O(\frac{1}{N})$. Similarly, the belief-action term $\mathcal{T}_a$ strongly depends on a term, $\langle q \rangle (1-\langle p \rangle) + \langle p \rangle (1-\langle q \rangle)$, that reflects how beliefs and actions are mixed and differ between each other (large when $\langle q \rangle, \langle p \rangle \simeq \frac{1}{2}$ or $|\langle q \rangle - \langle p \rangle|$ large). The contribution of the correlation between the individual's belief and action $Cov(q,p)$ is, analogously to the variance in $\mathcal{T}_b$, subleading, of order $O(\frac{1}{N})$.
\\
\\
\noindent \textbf{Imitation Dynamics.} 
Imitation dynamics in simulations proceed as follows.
\begin{enumerate}
    \item The population fitness of each player $i$ is calculated as $\pi_i=\frac{1}{N-1}\sum_{j\neq i}^N\pi_{ij}$ where $N$ is the population size.
    \item A pair of players, $k$ and $l$ are chosen at random (with the constraint $k\neq l$. Player $k$ then adopts the strategy fo player $l$ with probability $\Pi_{k\to l}=\frac{1}{1+\exp[s(\pi_k-\pi_l)]}$. This imitation event involves copying all of player $l$'s traits (i.e. any traits that are subject to evolution, including $p$, $q$, $\tau$, $\psi$ ir $\theta$). The parameter $s$ is the selection intensity.
    \item Regardless of whether $k$ copied $l$'s strategy, each of $k$'s evolving traits are independently subject to random mutation with probability $\mu$. The size of the mutation is drawn uniformly from the interval $\pm\Delta \mu$ around the current trait value, subject to boundary conditions such that all traits must lie in $[0,1]$.
    \end{enumerate}


\newpage

\begin{centering}
{\huge
\textbf{Supplementary Information}
}
\end{centering}

\tableofcontents
\newpage

\setcounter{equation}{9}

In this supplement we provide details of the analysis of the evolutionary model and robustness checks of our findings through simulation.

\section{Normal-form game with belief}

We focus on a classic $2\times2$ normal-form game, in which players choose to cooperate ($CO$) or defect ($DE$) such that cooperating entails a cost $c>0$ and generates a benefit $b>0$, where if both players cooperate, they share the cost, and the whole benefit is gained if either player cooperates. If $b>c$ this is a Snowdrift game, and if $c/2<b<c$ it is a Prisoner's Dilemma. 

Integrating belief into this game generates a $4\times4$ payoff matrix, with composite actions $(CO,co)$ (cooperate and express belief in cooperation), $(CO,de)$ (cooperate and express belief in defection), $(DE,co)$ (defect and express belief in cooperation) and $(DE,de)$ (defect and express belief in defection) -- as described in the main text and shown in Figure S4). 

\begin{figure*}[!h]
\centering
\includegraphics[width=1\textwidth]{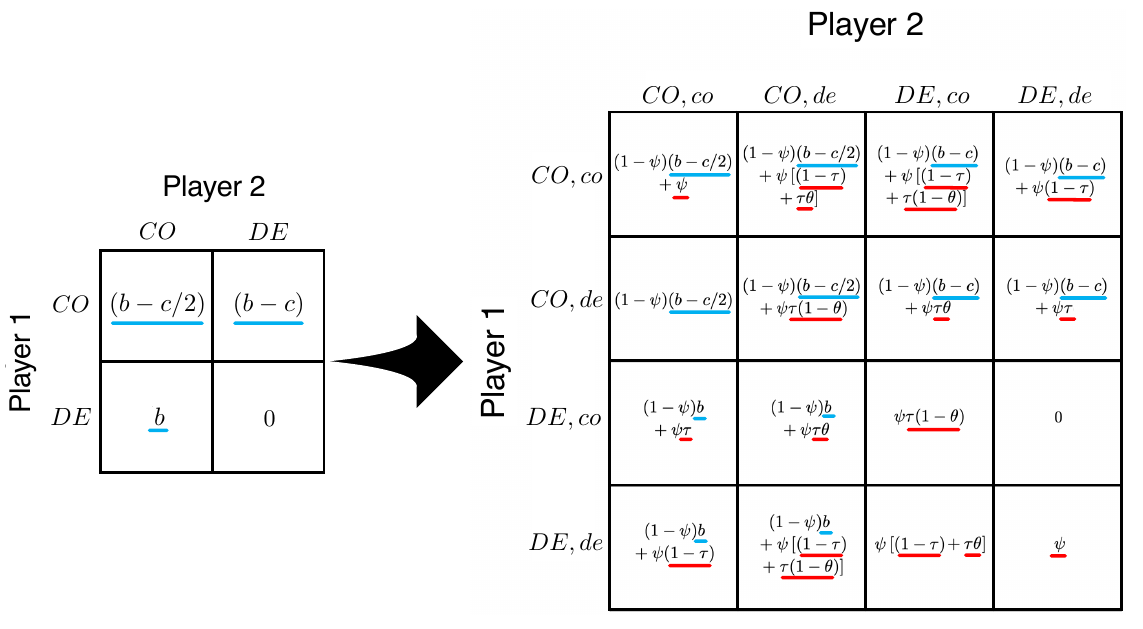}
\caption*{\textbf{Figure S1 - Transformed payoff matrix.} }\end{figure*}

The (untransformed) normal-form game with a $2\times2$ payoff matrix has been widely studied and has a single Nash equilibrium -- the pure strategy $(DE,DE)$ i.e. always defect -- when the game is a Prisoner's dilemma ($c/2<b<c$), and three Nash equilibria  -- corresponding to two pure strategies $(CO,DE)$, $(DE,CO)$ and a mixed strategy $\left(\frac{2(b-c)}{2b-c},\frac{2(b-c)}{2b-c}\right)$ i.e. either one player always cooperates and the other always defects, or else both players cooperate with probability $\frac{2(b-c)}{2b-c}$ -- when it is a Snowdrift game ($b>c$). 
Below, we calculate the pure strategy Nash equilibria for the transformed $4\times4$ payoff matrix.

\subsection{Pure strategy Nash equilibria}

To calculate the pure strategy Nash equilibria for the transformed $4\times4$ payoff matrix (Figure S1) we calculate the best response for each of the pure strategies $(X_i,x_i)$. In general, this depends on the values of $\psi$, $\theta$, and $\tau$ as well as the costs and benefits of cooperation, $b$ and $c$. This yields six possible Nash equilibria as follows:
\\
\\
\textbf{Case I: Cooperate and believe in cooperation.}
The pure strategy $(CO,co)$ is a best response to itself provided

$$
2\frac{\psi\tau}{1-\psi}>c
$$

and

$$
2\frac{\psi(1-\tau)}{1-\psi}>c
$$

If the first condition is violated both players are incentivised to switch to $(DE,co)$ and if the second condition is violated both players are incentivised to switch to $(DE,de)$.

When these two conditions are met, the pure strategy profile $\{(CO,co),(CO,co)\}$ is a Nash equilibrium. 
\\
\\
\textbf{Case II: Defect and believe in defection.}
The pure strategy $(DE,de)$ is a best response to itself provided

$$
\frac{\psi\tau}{1-\psi}>b-c
$$

and

$$
\frac{\psi(1-\tau)}{1-\psi}>b-c
$$

If the first condition is violated both players are incentivised to switch to $(CO,co)$ and if the second condition is violated both players are incentivised to switch to $(CO,de)$.

When these two conditions are met, the pure strategy profile $\{(DE,de),(DE,de)\}$ is a Nash equilibrium. Note that this condition is \emph{always} met when $b<c$ (i.e. the game is a Prisoner's Dilemma)
\\
\\
\textbf{Case III: Defect and believe in cooperation.}
The pure strategy $(DE,co)$ is a best response to itself provided

$$
 \frac{\psi (1 - \tau)}{(1 - \psi)}<- (b-c) 
$$

and

$$
\frac{\psi \tau}{1-\psi}>b-c
$$

and

$$
\tau>\frac{1}{2}
$$

If the first condition is violated both players are incentivised to switch to $(CO,co)$, if the second condition is violated both players are incentivised to switch to $(CO,de)$ and if the third condition is violated both players are incentivized to switch to $(DE,de)$

When these two conditions are met, the pure strategy profile $\{(DE,co),(DE,co)\}$ is a Nash equilibrium. 
\\
\\
\textbf{Case IV: Believe in cooperation (mixed material actions).}
The pure strategies $(CO,co)$ and $(DE,co)$ are best responses to each other provided

$$
\frac{\psi \tau}{1-\psi}>-(b-c)
$$

and

$$
\frac{\psi(1-\tau)}{1-\psi}>-(b-c)
$$

and

$$
2\frac{\psi(1-\tau)}{1-\psi}<c
$$

and

$$
\tau>\frac{1}{2}
$$

If the first condition is violated the $(CO,co)$ player is incentivised to switch to $(DE,co)$, if the second condition is violated the $(CO,co)$ player is incentivised to switch to $(DE,de)$.
If the third condition is violated the $(DE,co)$ player is incentivised to switch to $(CO,co)$, and if the fourth condition is violated the $(DE,co)$ player is incentivised to switch to $(DE,de)$.

When all four conditions are met, the pure strategy profile $\{(CO,co),(DE,co)\}$ is a Nash equilibrium. 
\\
\\
\textbf{Case V: Believe in defection (mixed material actions).}
The pure strategies $(CO,de)$ and $(DE,de)$ are best responses to each other provided

$$
\tau>\frac{1}{2}
$$

and

$$
\frac{\psi \tau}{1-\psi}>-(b-c)
$$

and

$$
\frac{\psi(1-\tau)}{1-\psi}<b-c
$$

and

$$
2\frac{\psi(1-\tau)}{1-\psi}<c
$$

If the first condition is violated the $(CO,de)$ player is incentivised to switch to $(CO,co)$, if the second condition is violated the $(CO,de)$ player is incentivised to switch to $(DE,co)$ and if the third condition is violated the $(CO,de)$ player is incentivised to switch to $(DE,de)$. Finally if the fourth condition is violated the $(DE,de)$ player is incentivised to switch to $(DE,co)$.

When all four conditions are met, the pure strategy profile $\{(CO,de),(DE,de)\}$ is a Nash equilibrium. 
\\
\\
\textbf{Case VI: Mixed beliefs and material actions.}
The pure strategies $(CO,co)$ and $(DE,de)$ are best responses to each other provided

$$
\frac{\psi(1 - \tau)}{ (1 - \psi) }>-(b - c)
$$

and

$$
\tau<\frac{1}{2}
$$

and

$$
\frac{\psi \tau}{ (1 - \psi) }<(b - c)
$$

and

$$
2\frac{\psi \tau}{1-\psi}<c
$$

If the first condition is violated the $(CO,co)$ player is incentivised to switch to $(DE,co)$, if the second condition is violated the $(CO,co)$ player is incentivised to switch to $(CO,de)$ and the $(DE,de)$ player is incentivised to switch to $(DE,co)$, and if the third condition is violated the $(CO,co)$ player is incentivised to switch to $(DE,de)$. Finally if the fourth condition is violated the $(DE,de)$ player is incentivised to switch to $(CO,co)$.

When all four conditions are met, the pure strategy profile $\{(CO,co),(DE,de)\}$ is a Nash equilibrium.

\subsubsection{Qualitative description of Nash equilibria}

We find six possible pure strategy Nash equilibria, which include cases where beliefs and material actions are misaligned for one or both players (Case III, IV and V), cases where material actions differ between the players (Case IV, V and VI) and a case where beliefs differ between the players (Case VI). 

We note that none of the conditions for the six strategy profiles to be Nash equilibria depend on the weight given to injunctive vs descriptive norms, $\theta$, but do depend on the relative weight of material vs psychological payoff $\psi$ and on the weight given to meta-norms of conformity vs coherence $\tau$, as well as the material payoffs of the game, $b$ and $c$. This leads to different Nash equilibria existing in the Snowdrift game vs the Prisoner's Dilemma (Figure S2), although the qualitative effect of varying the parameters $\psi$ and $\tau$ is similar in both cases.

Next we analyze the evolutionary dynamics of belief, behavior and meta-norms associated with the transformed game.

\begin{figure*}[!h]
\centering
\includegraphics[width=1\textwidth]{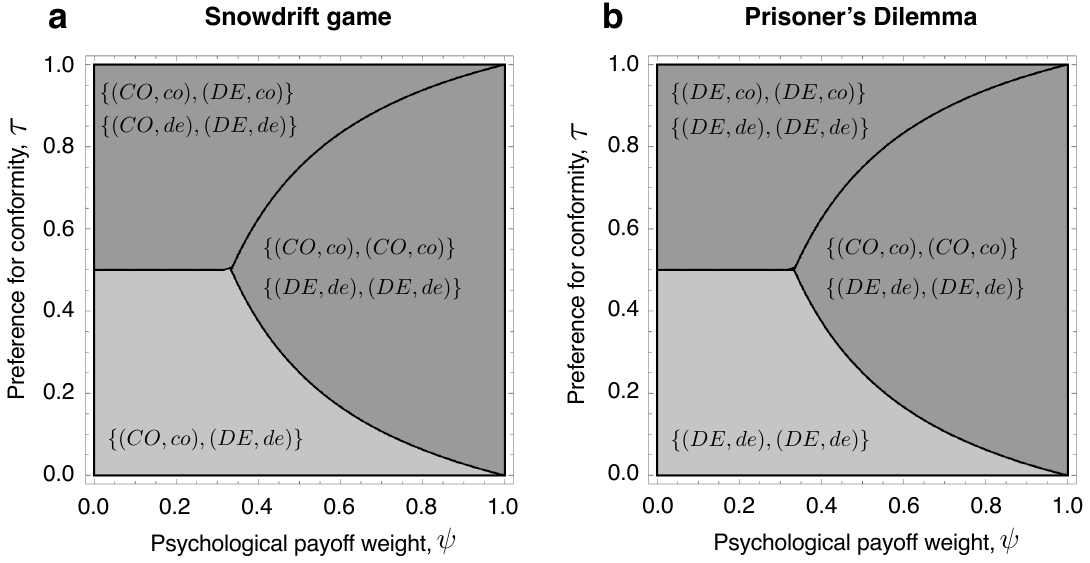}
\caption*{\textbf{Figure S2 - Pure strategy Nash equilibria for the transformed $4\times 4$ payoff matrix.} }\end{figure*}

\clearpage

\section{Analysis of evolutionary dynamics}

We study the evolutionary dynamics of behavior, belief and meta-norms using a combination of evolutionary invasion analysis and simulations.  This allows us to account for mixed strategies, i.e. situations in which an individual may used different actions, and/or express different belief in different interactions. In this section we describe the details of evolutionary invasion analysis, as well as several extensions to the results presented in the main text. Note that in the sections below, the locations of equilibria and the form of the selection gradient are calculated using Mathematica, and the file can be found at the Github repository for this paper.

\subsection{Two-trait model}

As described in the main text, we make the simplifying assumption that action and belief correspond to two ``traits'',  $p$ and $q$, where $p$ is the probability that, in a given interaction, a player uses the action cooperate and $q$ is the probability that they express belief in cooperation. Thus the probability that the players uses composite action $(CO,co)$ is $pq$ and so on.

We begin by determining the evolutionary singular strategies of the system, that is, the trait values at which a population is monomorphic, and at which selection does not favor invasion by any alternative ``mutant'' strategy that is similar to the current, resident strategy. That is, we look at the stability of a resident strategy $\{p_r,q_r\}$ in a population against an invader $i$ with strategy $\{p_i,q_i\}$. 

In general, the evolution of a trait vector $\mathbf{x}$ describing the resident strategy for the population, with invasion occurring via local mutations, proceeds according to

\begin{equation}
\frac{d \mathbf{x}}{dt}=N(\mathbf{x})\mu\Sigma(\mathbf{x})\mathbf{s}(\mathbf{x})
\end{equation}

where $N$ is the effective population size, $\mu$ is the mutation rate per imitation/birth event and $\Sigma$ is the mutational variance– covariance matrix summarizing the distribution of mutations around the current trait value. The vector $\mathbf{s}(\mathbf{x})$ gives the selection gradient of each trait $k$ i.e. $s_k(\mathbf{x})=\frac{\partial \pi(\mathbf{x})}{\partial x_k}=0$ (where $\pi(\mathbf{x})$ is the utility function).
Under our model we will assume these distributions to be constant except at the boundary (where the constraint that probabilities lie in $[0,1]$ prevents mutations that decrease or increase $p$ and $q$). We also assume that $N$ and $\mu$ are constant.

The evolutionary dynamics have an equilibrium if the left hand side of Eq. 10 is zero. If the trait value does not lie at the boundary, this can only occur if the selection gradient $\mathbf{s}(\mathbf{x})=0$, given our assumption that $\mu$, $N$ and $\Sigma$ are constant. If the trait value does lie at the boundary, no mutations can occur that move the trait value beyond the boundary. Thus the evolutionary dynamics can have an equilibrium at the boundary provided the selection gradient is orthogonal and into the boundary (since the component of the $\Sigma$ in the direction of the selection gradient is zero). For example, if $q_r=1$ and $s_q(\mathbf{x})>0$, such that selection acts to increase $q$, and assuming the selection gradient is either zero or orthogonal and into the boundary for all other traits, then an equilibrium is reached.

For equilibria at which the selection gradient is zero for all traits, stability is determined by assessing whether they are convergent stable, and whether they are subject to diversifying selection (see below).
For equilibria in which some traits have zero selection gradient, while others reach their physical max/min, stability analysis (i.e. determining whether there is convergence stability and diversifying selection) is required for those traits with zero selection gradient.

\subsubsection{Utility function for the two-trait model}

We first consider a two-trait model, in which an individual $i$ is characterized only by their probability of cooperation, $p_i$, and their probability of expressing belief in cooperation, $q_i$. We initially set all other parameters, $\psi$, $\tau$, $\theta$, $b$ and $c$ to be constant and the same for all players. 

The utility for a player $i$ interacting with a player $j$ under this model is

\begin{align}
\nonumber \pi_{ij}=p_iq_i
\Big[p_jq_j((1-\psi)(b - c/2) + \psi) + (1 - p_j)
      q_j((1-\psi)(b - c) + \psi(1-\tau) + (1 - \theta)\psi\tau ) \\
\nonumber      + p_j(1 - q_j)((1-\psi)(b - c/2) + \psi(1-\tau) + \psi\theta \tau ) + (1 - 
        p_j)(1 - q_j)((1-\psi)(b - c) + \psi(1-\tau))\Big] \\
 \nonumber       + (1 - p_i)
     q_i\Big[p_jq_j((1-\psi)b +  \psi\tau + (1 - p_j)
      q_j(\psi(1 - \theta)\tau ) 
         p_j(1 - q_j)((1-\psi)b + \psi\theta \tau \Big] \\
   \nonumber     + p_i(1 - q_i)\Big[p_j
      q_j(1-\psi)(b - c/2) + (1 - p_j)
      q_j((1-\psi)(b - c) + \psi\theta \tau) \\
  \nonumber    + p_j(1 - q_j)((1-\psi)(b - c/2) + \psi(1 - \theta)\tau ) + (1 - 
        p_j)(1 - q_j)((1-\psi)(b - c) + \psi\tau )\Big] \\
    \nonumber    + (1 - 
       p_i)(1 - q_i)\Big[p_jq_j((1-\psi)b + \psi(1-\tau)) + (1 - p_j)
      q_j(\psi(1-\tau) + \psi\theta \tau)\\
      + p_j(1 - q_j)((1-\psi)b + \psi(1-\tau) + \psi(1 - \theta)\tau) + (1 - p_j)(1 - 
        q_j)(\psi(1-\tau) +  \psi\tau)\Big]
\end{align}

 We assume that the selection acts on the total utility derived from engaging in all pairwise interactions between the focal player $i$ and all other members of the population, $\pi_i=\frac{1}{N-1}\sum_{j\neq i}\pi_{ij}$ (as described in the main text). 

In order to determine the equilibria of Eq. 10 we first calculate the selection gradient for each trait, $\frac{\partial \pi}{\partial q_r}$ and $\frac{\partial \pi}{\partial p_r}$, where $p_r$ and $q_r$ is the resident strategy of the population, i.e. such that $\pi_i=\pi_{ir}$.

\subsubsection{Evolutionary singular strategies}
We first calculate the conditions for an equilibrium away from the boundary, i.e. values of $(p^*_r,q^*_r)$ such that

$$
\frac{\partial \pi_i}{\partial q_i}\Big |_{q_i=q_r}=\frac{\partial \pi_i}{\partial p_i}\Big |_{p_i=p_r}=0
$$

Calculating the derivatives we find that such a solution must satisfy

\begin{eqnarray}
\nonumber q_r^*&=& p_r^* +\frac{1 - 2p_r^*}{\tau (1 - \theta )}\\
\nonumber p_r^*&=&\frac{2 (b - c) }{2 b - c} - \frac{2 \psi (1 - \tau)}{(1 - \psi)(2 b - c)} (1 - 2 q^*_r )\\
\end{eqnarray}

which may in general produce a viable strategy $p_r^*\in[0,1]$ and $q_r^*\in[0,1]$. However to determine whether such equilibrium is convergence stable \cite{Mullon:2016aa} we must calculate the eigenvalues of the Jacobian matrix $\mathbf{J}(p,q)$, with entries

$$
\mathbf{J}(p,q)=\begin{pmatrix}
\frac{\partial \mathbf{s}_p}{\partial p}  & \frac{\partial \mathbf{s}_p}{\partial q}\\
\frac{\partial \mathbf{s}_q}{\partial p}  & \frac{\partial \mathbf{s}_q}{\partial q}
\end{pmatrix}
$$

where $\mathbf{s}_q=\frac{\partial \pi_i}{\partial q_i}$ and $\mathbf{s}_p=\frac{\partial \pi_i}{\partial p_i}$ are the selection gradients of the two traits. The equilibrium of $\mathbf{J}$ have negative real part, and Eq. 12 is stable provided $\text{tr}(\mathbf{J})<0$ and $\text{det}(\mathbf{J})>0$. Calculating the trace and determinant of $\mathbf{J}$ at $(p_r^*,q_r^*)$ we find

\begin{eqnarray}
\nonumber    \text{tr}(\mathbf{J})&=&2 \psi (1 - \theta) \tau-\frac{1}{2} (2 b - c) (1 - \psi)\\
\nonumber     \text{det}(\mathbf{J})&=&-\psi(1 - \psi)(2 b - c)(1 - \theta)\tau - 
 4\psi ^2((1 - \tau) + \theta\tau)(1 - \tau)\\
\end{eqnarray}

Since $2b>c$ is required for the game to be a Prisoner's Dilemma or a Snowdrift game, and the norm utility weights $\psi$, $\tau$ and $\theta$ all lie in $[0,1]$, $\text{det}(\mathbf{J})<0$ and the equilibrium is unstable. In addition, note that if $2b<c$, $\text{tr}(\mathbf{J})>0$ and the equilibrium is again unstable. 

Thus we conclude that no stable equilibrium exists in the two-trait model such that both $p$ and $q$ both lie in the interior of the strategy space.

\subsubsection{Equilibria at the boundary}

Next we consider the stability of equilibria in which one or both of the traits lie at the boundary of the strategy space. This yields eight cases which we analyze in turn.
\\
\\
\noindent \textbf{Case I: $q_r^*=0$ and $0<p_r^*<1$.}
Setting $q^*_r=0$, Eq. 12 yields

$$
p_r^*=\frac{2 (b - c) }{2 b - c} - \frac{2 \psi (1 - \tau)}{(1 - \psi)(2 b - c)}
$$
which is a viable strategy provided

$$
\frac{\psi(1-\tau)}{1-\psi}<b - c
$$
and $b>c$.
We first look at the selection gradient of $q$ and we require $\mathbf{s}_q(p_r^*,0)\leq 0$ for stability. This yields the condition

$$
 4\frac{\psi(1 - \tau)}{1 - \psi}   \geq \frac{2(b - c )(1-2(1-\theta)\tau)-c } {1-(1-\theta)\tau} 
$$

If $\theta=0$ (injunctive norms) this simplifies to

$$
 4\frac{\psi}{1 - \psi}   \geq \frac{2(b - c )(1-2\tau)-c } {(1-\tau)^2} 
$$
 whereas if $\theta=1$ (descriptive norms) we recover

$$
 4\frac{\psi}{1 - \psi}   \geq \frac{2b - 3c} {1-\tau} 
$$
 both of which conditions can be satisfied for viable choices of $\tau$, $\psi$, $b$ and $c$ (see Figure S3). 

Next we consider the stability of the equilibrium point at $p_r^*$ when $q_r^*=0$. The system is reduced to one dimension and stability requires $\frac{\partial \mathbf{s}_p}{\partial p}\Big |_{p=p_r^*}<0$. This yields the condition
$$
(2 b - c) (1 - \psi)\geq0
$$
which is satisfied for both the Prisoner's Dilemma and the Snowdrift game provided $\psi>0$. And so the equilibrium with $q_r^*=0$ and $0<p_r^*<1$ is conditionally stable (see Figure S3).
 \\
 \\
\noindent \textbf{Case II: $q_r^*=1$ and $0<p_r^*<1$.}
Setting $q^*_r=1$, Eq. 12 yields

$$
p_r^*=\frac{2 (b - c) }{2 b - c} + \frac{2 \psi (1 - \tau)}{(1 - \psi)(2 b - c)}
$$
which is a viable strategy provided

$$
 2 \frac{\psi (1 - \tau)}{1-\psi}<c
$$
and $b>c$.
We first look at the selection gradient of $q$ and we require $\mathbf{s}_q(p_r^*,1)\geq 0$ for stability. This yields the condition

$$
 4\frac{\psi(1 - \tau)}{1 - \psi}   \geq \frac{c(1-2(1-\theta)\tau)-2(b-c) } {1-(1-\theta)\tau} 
$$

If $\theta=0$ (injunctive norms), this equation simplifies to

$$
  4\frac{\psi}{1 - \psi}   \geq \frac{c(1-2\tau)-2(b-c) } {(1-\tau)^2} 
$$
 whereas if $\theta=1$ (descriptive norms), we recover
$$
  4\frac{\psi}{1 - \psi}   \geq \frac{3c-2b } {1-\tau}  
$$
 Both conditions can be satisfied for viable choices of $\tau$, $\psi$, $b$ and $c$ (see Figure S3). 

Next we consider the stability of the equilibrium point at $p_r^*$ when $q_r^*=1$. The system is reduced to 1-D and stability requires $\frac{\partial \mathbf{s}_p}{\partial p}\Big |_{p=p_r^*}<0$. This again yields the condition

$$
(2 b - c) (1 - \psi)>0
$$
which is satisfied for both the Prisoner's Dilemma and the Snowdrift game provided $\psi>0$. And so the equilibrium with $q_r^*=1$ and $0<p_r^*<1$ is conditionally stable (see Figure S3).
\\
 \\
\noindent \textbf{Case III: $0<q_r^*<1$ and $p_r^*=0$.}
Setting $p^*_r=0$, Eq. 12 yields

$$
q_r^*=\frac{1}{\tau (1 - \theta )}
$$
which is not a viable strategy given the constraints on $\tau$ and $\theta$ (except in the limit $\tau=1$ and $\theta=0$, in which case $q_r^*=1$, which we discuss below). And so an equilibrium with $p_r^*=0$ and $0<q_r^*<1$ is not possible.
\\
 \\
\noindent \textbf{Case IV: $0<q_r^*<1$ and $p_r^*=1$.}
Setting $p^*_r=1$, Eq. 12 yields

$$
q_r^*=1 -\frac{1}{\tau (1 - \theta )}
$$
which is not a viable strategy given the constraints on $\tau$ and $\theta$ (except in the limit $\tau=1$ and $\theta=0$, in which case $q_r^*=0$, which we discuss below). An equilibrium with $p_r^*=0$ and $0<q_r^*<1$ is therefore not possible.
\\
 \\
\noindent \textbf{Case V: $p_r^*=0$ and $q_r^*=0$.}
In this case we require $\mathbf{s}_q(0,0)\leq 0$ and  $\mathbf{s}_p(0,0)\leq 0$ for stability. Calculating the selection gradients we recover the conditions

$$
\frac{\psi (1 - \tau)}{1 - \psi}\geq b - c
$$
and $\psi\geq0$. And so the equilibrium with $p_r^*=0$ and $q_r^*=0$ is conditionally stable (see Figure S3).
\\
 \\
\noindent \textbf{Case VI: $p_r^*=1$ and $q_r^*=0$.}
In this case we require $\mathbf{s}_q(1,0)\leq 0$ and  $\mathbf{s}_p(1,0)\geq 0$ for stability. Calculating the selection gradients we recover the conditions

$$
2\frac{\psi (1 - \tau)}{1-\psi}\leq -c
$$
and 

$$
1 - 2  (1 - \theta) \tau\leq0
$$
The first condition cannot be satisfied.
And so there is no stable equilibrium with $p_r^*=1$ and $q_r^*=0$ (see Figure S3).
\\
 \\
\noindent \textbf{Case VII: $p_r^*=0$ and $q_r^*=1$.}
In this case we require $\mathbf{s}_q(0,1)\geq 0$ and  $\mathbf{s}_p(0,1)\leq 0$ for stability. Calculating the selection gradients we recover the conditions

$$
\frac{\psi(1 - \tau)}{1-\psi} \leq -(b - c)
$$
which can be satisfied if $b<c$ and 

$$
 1 - 2 (1 - \theta) \tau \leq 0
$$
which can be satisfied for injunctive norms ($\theta=0$) but not descriptive norms.
And so the equilibrium with $p_r^*=1$ and $q_r^*=0$ is conditionally stable (see Figure S3).
\\
\\
\noindent \textbf{Case VIII: $p_r^*=1$ and $q_r^*=1$.}
In this case we require $\mathbf{s}_q(1,1)\geq 0$ and  $\mathbf{s}_p(1,1)\geq 0$ for stability. Calculating the selection gradients we recover the conditions

$$
2\frac{\psi(1 - \tau)}{1-\psi}\geq c
$$
and $\psi\geq 0$
And so the equilibrium with $p_r^*=1$ and $q_r^*=1$ is conditionally stable (see Figure S3).

\subsubsection{Qualitative description of Nash equilibria}

We find five possible stable equilibria of the evolutionary dynamics, which include cases where beliefs and material actions are misaligned for one or both players (Case 1, II and VII), cases where material actions can differ between the players in a given interaction (Case I) but we do not find cases where variation in belief across individuals is stable. 

In contrast to the pure strategy Nash equilibria, we note that the conditions for stability may depend on the weight given to injunctive vs descriptive norms, $\theta$ (Case VII in particular, see Figure S3) as well as the relative weight of material vs psychological payoff $\psi$ and on the weight given to meta-norms of conformity vs coherence $\tau$, as well as the material payoffs of the game, $b$ and $c$. This leads to the population converging on different equilibria in the Snowdrift game vs the Prisoner's Dilemma, although the qualitative effect of varying the parameters $\psi$ and $\tau$ is similar in both cases.

\begin{figure*}[!h]
\centering
\includegraphics[width=1\textwidth]{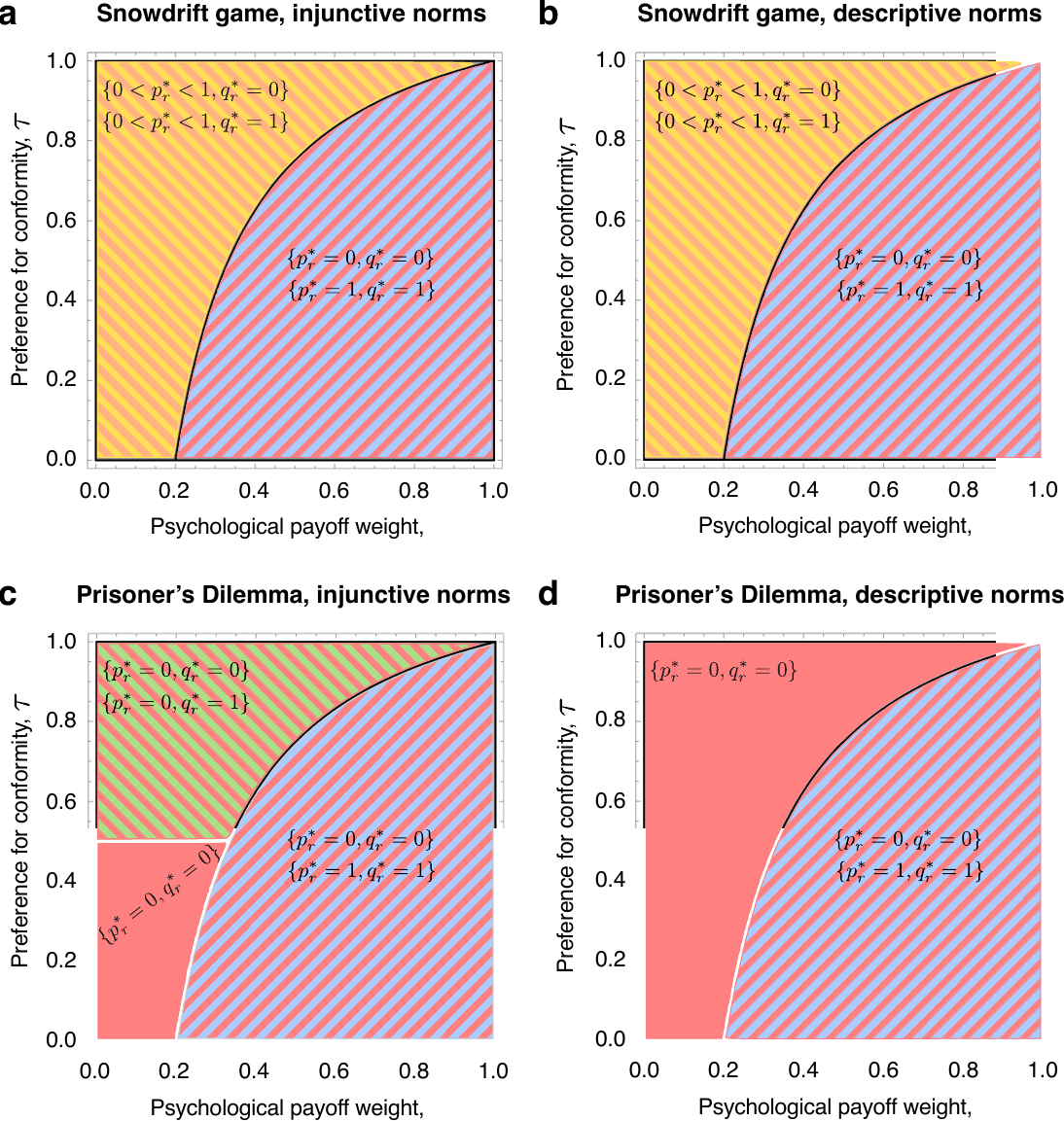}
\caption*{\textbf{Figure S3 - Stable equilibria for the two-trait model, under evolutionary invasion analysis.} }\end{figure*}

\clearpage

\subsubsection{Branching}

Finally we consider whether the two-trait model can support evolutionary branching. Qualitatively, branching will occur when a convergent stable equilibrium experiences diversifying selection, leading to the population supporting sustained polymorphism (i.e. the coexistence over time of distinct ``groups'' who share similar traits with in-group members, but are diverged from out-group members).

Because diversifying selection arises from interactions between rare mutants, it is a second order effect, and is only expected to arise at points of zero selection gradient. Thus the only viable candidates for branching are Case I and Case II above, occurring with respect to $p$, the probability of cooperation.

The effect of interactions between mutants is captured by $\mathbf{H}(p,q)$ which has entries

$$
\mathbf{H}(p,q)=\begin{pmatrix}
\frac{\partial^2 \pi_i}{\partial p_i^2}  & \frac{\partial^2 \pi_i}{\partial p_i\partial q_i}\\
\frac{\partial^2 \pi_i}{\partial q_i\partial p_i}  & \frac{\partial^2 \pi_i}{\partial q_i^2}
\end{pmatrix}
$$
If $\mathbf{H}$ has positive eigenvalues, mutants impact utility synergistically, and branching can occur \cite{Mullon:2016aa}. In Case I and Case II, we are only concerned with the 1-D system and thus the sign of $\frac{\partial^2 \pi_i}{\partial p_i^2}\Big |_{p_i=p_r^*}$. We find

$$
\frac{\partial^2 \pi_i}{\partial p_i^2}\Big |_{p_i=p_r^*}=0
$$

and so we do not expect selection to favor branching.

\clearpage

\subsection{Three-trait models}

In this section we analyze three-trait models, in which $p$ and $q$, as well as one of $\tau$, $\psi$ or $\theta$ are allowed to evolve.

\subsubsection{Three-trait model with evolving $\tau$}

Next we consider a three-trait model, in which an individual $i$ is characterized by their probability of cooperation, $p_i$, their probability of expressing belief in cooperation, $q_i$ and their preference for conformity, $\tau_i$. We set all other parameters, $\psi$, $\theta$, $b$ and $c$ to be constant and the same for all players. 

The utility for a player $i$ interacting with a player $j$ under this model is

\begin{align}
\nonumber \pi_{ij}=p_iq_i
\Big[p_jq_j((1-\psi)(b - c/2) + \psi) + (1 - p_j)
      q_j((1-\psi)(b - c) + \psi(1-\tau_i) + (1 - \theta)\psi\tau_i ) \\
\nonumber      + p_j(1 - q_j)((1-\psi)(b - c/2) + \psi(1-\tau_i) + \psi\theta \tau_i ) + (1 - 
        p_j)(1 - q_j)((1-\psi)(b - c) + \psi(1-\tau_i))\Big] \\
 \nonumber       + (1 - p_i)
     q_i\Big[p_jq_j((1-\psi)b +  \psi\tau_i + (1 - p_j)
      q_j(\psi(1 - \theta)\tau_i ) 
         p_j(1 - q_j)((1-\psi)b + \psi\theta \tau_i \Big] \\
   \nonumber     + p_i(1 - q_i)\Big[p_j
      q_j(1-\psi)(b - c/2) + (1 - p_j)
      q_j((1-\psi)(b - c) + \psi\theta \tau_i) \\
  \nonumber    + p_j(1 - q_j)((1-\psi)(b - c/2) + \psi(1 - \theta)\tau_i ) + (1 - 
        p_j)(1 - q_j)((1-\psi)(b - c) + \psi\tau_i )\Big] \\
    \nonumber    + (1 - 
       p_i)(1 - q_i)\Big[p_jq_j((1-\psi)b + \psi(1-\tau_i)) + (1 - p_j)
      q_j(\psi(1-\tau_i) + \psi\theta \tau_i)\\
      + p_j(1 - q_j)((1-\psi)b + \psi(1-\tau_i) + \psi(1 - \theta)\tau_i) + (1 - p_j)(1 - 
        q_j)(\psi(1-\tau_i) +  \psi\tau_i)\Big]
\end{align}
which is identical to Eq. 11, with $\tau=\tau_i$.

\subsubsection{Selection on $\tau$}
In order to identify the stable the equilibria of the three-trait model, we first look for evolutionary singular strategies, which satisfy

$$
\frac{\partial \pi_i}{\partial q_i}\Big |_{q_i=q_r}=\frac{\partial \pi_i}{\partial p_i}\Big |_{p_i=p_r}=\frac{\partial \pi_i}{\partial \tau_i}\Big |_{\tau_i=\tau_r}=0
$$
This yields a single solution

\begin{eqnarray}
\nonumber q_r^*&=&\frac{1}{2}\\
\nonumber p_r^*&=&\frac{2 (b - c) }{2 b - c} \\
\nonumber \tau_r^*&=&\frac{1}{1-\theta}\\
\end{eqnarray}

This only produces a physical solution in the limiting case $\theta=0$ (purely injunctive norms), in which case $\tau_r^*=1$, thus it falls within the cases discussed below. Specifically, we look at equilibria for which $\tau_r^*=0$, or $\tau_r^*=1$, and then leverage the results above to determine the stability of $p_r^*$ and $q_r^*$. 

Looking in detail at the selection gradient associated with $\tau_i$ we find

$$
\frac{\partial \pi_i}{\partial \tau_i}\Big |_{\tau_i=\tau_r}=\psi (p_r^* - q_r^*) (1 - 2 q_r^*) (1 - \theta)
$$
which we note is independent of $\tau_r^*$. We then have three possible scenarios for the evolution of $\tau$. 
\\
\\
\noindent \textbf{Case I: $\frac{\partial \pi_i}{\partial \tau_i}\Big |_{\tau_i=\tau_r}<0$.}
If $\frac{\partial \pi_i}{\partial \tau_i}\Big |_{\tau_i=\tau_r}<0$ at equilibrium, then the only viable equilibrium occurs at $\tau_r^*=0$. Such an equilibrium will arise when $\psi>0$, $\theta<1$ and when $p_r^*<q_r^*$ and $q_r^*<1/2$ or $p_r^*>q_r^*$ and $q_r^*>1/2$.

We can use the stability conditions for the two-trait model, with $\tau=0$, to determine whether such an equilibrium can be stable. There are five possible cases as described in the previous section. Taking in these in turn we find

\begin{enumerate}
\item Taking $q_r^*=0$ and $0<p_r^*<1$, results in $\frac{\partial \pi_i}{\partial \tau_i}\Big |_{\tau_i=\tau_r}>0$ and so cannot be stable.
\item Taking $q_r^*=1$ and $0<p_r^*<1$, results in $\frac{\partial \pi_i}{\partial \tau_i}\Big |_{\tau_i=\tau_r}>0$ and so cannot be stable.
\item Taking $q_r^*=0$ and $p_r^*=0$, results in $\frac{\partial \pi_i}{\partial \tau_i}\Big |_{\tau_i=\tau_r}=0$ for all $\tau_r^*$ (see below).
\item Taking $q_r^*=1$ and $p_r^*=0$, results in $\frac{\partial \pi_i}{\partial \tau_i}\Big |_{\tau_i=\tau_r}>0$ and so cannot be stable.
\item Taking $q_r^*=1$ and $p_r^*=1$, results in $\frac{\partial \pi_i}{\partial \tau_i}\Big |_{\tau_i=\tau_r}=0$ or all $\tau_r^*$ (see below).
\end{enumerate}

And so we see there is no scenario in which $\tau_r^*=0$ is a non-degenerate stable equilibrium.
\\
\\
\noindent \textbf{Case II: $\frac{\partial \pi_i}{\partial \tau_i}\Big |_{\tau_i=\tau_r}>0$.}
If $\frac{\partial \pi_i}{\partial \tau_i}\Big |_{\tau_i=\tau_r}>0$ at equilibrium, then the only viable equilibrium occurs at $\tau_r^*=1$. Such an equilibrium will arise when $\psi>0$, $\theta<1$ and when $p_r^*<q_r^*$ and $q_r^*>1/2$ or $p_r^*>q_r^*$ and $q_r^*<1/2$.

We can use the stability conditions for the two-trait model, with $\tau=1$, to determine whether such an equilibrium can be stable. Once again, there are five possible cases as described in the previous section. Taking in these in turn we find

\begin{enumerate}
\item Taking $q_r^*=0$ and $0<p_r^*<1$, results in $\frac{\partial \pi_i}{\partial \tau_i}\Big |_{\tau_i=\tau_r}>0$. Applying the relevant stability condition (Case I from Section 2.1.3) with $\tau=1$ we require $b - c>0$ and $ \frac{2b-c}{4(b-c)} \geq \theta$.
\item Taking $q_r^*=1$ and $0<p_r^*<1$, results in $\frac{\partial \pi_i}{\partial \tau_i}\Big |_{\tau_i=\tau_r}>0$. Applying the relevant stability condition (Case II from Section 2.1.3) with $\tau=1$ we require $c>0$ and $ \frac{2b-c}{2c} \geq \theta$.
\item Taking $q_r^*=0$ and $p_r^*=0$, results in $\frac{\partial \pi_i}{\partial \tau_i}\Big |_{\tau_i=\tau_r}=0$ for all $\tau_r^*$ (see below).
\item Taking $q_r^*=1$ and $p_r^*=0$, results in $\frac{\partial \pi_i}{\partial \tau_i}\Big |_{\tau_i=\tau_r}>0$. Applying the relevant stability condition (Case VII from Section 2.1.3) with $\tau=1$ we require $-(b-c)\geq0$ and $\theta \leq 1/2$.
\item Taking $q_r^*=1$ and $p_r^*=1$, results in $\frac{\partial \pi_i}{\partial \tau_i}\Big |_{\tau_i=\tau_r}=0$ or all $\tau_r^*$ (see below).
\end{enumerate}

And so we there are three distinct non-degenerate stable equilibria. with $\tau_r^*=1$.
\\
\\
\noindent \textbf{Case III: $\frac{\partial \pi_i}{\partial \tau_i}\Big |_{\tau_i=\tau_r}=0$.}
If either $\psi=0$, $\theta=1$, else for any equilibrium of the two-trait model such that $p_r^*=q_r^*$, then selection on $\tau$ vanishes, $\frac{\partial \pi_i}{\partial \tau_i}\Big |_{\tau_i=\tau_r}=0$ (note we ignore the case $q_r^*=1/2$ as this is not an equilibrium of the two trait model).

When this occurs $\tau$ is released from selection, i.e. all values of $\tau$ experience zero selection gradient. Under evolutionary dynamics in a finite population, this results in ``drift'' through neutral invasions, such that the trait value of $\tau$ undergoes a random walk. We discuss the effects of drift in simulations further in Section 3. When populations are very large $N\to\infty$, as assumed for evolutionary invasion analysis, neutral invasions do not occur, and the value of $\tau$ at equilibrium will depend on the choice of initial conditions.

If $\psi=0$ or $\theta=1$, the selection gradient for $\tau$ is zero for all $p_r$ and $q_r$, and so the equilibrium value of $\tau$ is simply given by the initial condition. If $\psi>0$  and $\theta<1$, and an equilibrium is reached such that $q_r^*=p_r^*=1$ or $q_r^*=p_r^*=0$, $\tau$ will experience selection before equilibrium is reached.

\subsubsection{Branching in the $p-\tau$ plane}

We can analyze the conditions for branching in the three-trait model by looking at the eigenvalues of

$$
\mathbf{H}(p,\tau)=\begin{pmatrix}
\frac{\partial^2 \pi_i}{\partial p_i^2}  & \frac{\partial^2 \pi_i}{\partial p_i\partial \tau_i}\\
\frac{\partial^2 \pi_i}{\partial \tau_i\partial p_i}  & \frac{\partial^2 \pi_i}{\partial \tau_i^2}
\end{pmatrix}
$$
Since diversifying selection is a second order effect, we must look  at equilibria with zero selection gradient for both $p$ and $\tau$ (since there are no non-boundary equilibria for $q$). This requires $\theta=1$ or $\psi=0$ (since the other equilibria of zero selection gradient for $\tau$ have non-zero selection gradient for $p$).

Calculating the eigenvalues of $\mathbf{H}(p,\tau)$ we find

$$
\lambda=\pm\psi(1 - 2q_r^*)
$$
and so there is always one negative and one positive eigenvalue when $\psi>0$ and $\theta=1$. Thus we expect evolutionary branching to occur at any stable equilibrium with $0<p_r^*<1$, when descriptive norms are used ($\theta=1$) and preference for conformity ($\tau$) is allowed to evolve.

\subsubsection{Qualitative description of Nash equilibria}

Taken together, we find six qualitative types of equilibrium for the three-trait model as follows

\begin{enumerate}
    \item $p_r^*=0$, $q_r^*=0$ corresponding to high social homogeneity (no branching), high tightness, low cooperation. At this equilibrium $\tau$ is released from selection.
    \item $p_r^*=1$, $q_r^*=1$ corresponding to high social homogeneity (no branching), high tightness, high cooperation. At this equilibrium $\tau$ is released from selection.
    \item $p_r^*<1$, $q_r^*=1$, $\theta<1$ corresponding to high social homogeneity (no branching), low tightness, low cooperation. At this equilibrium $\tau=1$.
    \item $p_r^*>0$, $q_r^*=0$, $\theta<1$ corresponding to high social homogeneity (no branching), low tightness, low cooperation. At this equilibrium $\tau=1$.
    \item $p_r^*<1$, $q_r^*=1$, $\theta=1$ corresponding to low social homogeneity (branching), low tightness, low cooperation. At this equilibrium $\tau$ and $p$ undergo branching.
    \item $p_r^*>0$, $q_r^*=0$, $\theta=1$ corresponding to low social homogeneity (branching), low tightness, low cooperation. At this equilibrium $\tau$ and $p$ undergo branching.
\end{enumerate}

These are the equilibria summarized in Figure 2 and 3 of the main text, and characterized via simulation in Figure S5-S6 and below.

\clearpage

\subsubsection{Three-trait model with evolving $\psi$}

Next we consider a three-trait model, in which an individual $i$ is characterized by their probability of cooperation, $p_i$, their probability of expressing belief in cooperation, $q_i$ and their preference for conformity, $\psi_i$. We set all other parameters, $\tau$, $\theta$, $b$ and $c$ to be constant and the same for all players. 

The utility for a player $i$ interacting with a player $j$ under this model is

\begin{align}
\nonumber \pi_{ij}=p_iq_i
\Big[p_jq_j((1-\psi_i)(b - c/2) + \psi_i) + (1 - p_j)
      q_j((1-\psi_i)(b - c) + \psi_i(1-\tau) + (1 - \theta)\psi_i\tau ) \\
\nonumber      + p_j(1 - q_j)((1-\psi_i)(b - c/2) + \psi_i(1-\tau) + \psi_i\theta \tau ) + (1 - 
        p_j)(1 - q_j)((1-\psi_i)(b - c) + \psi_i(1-\tau))\Big] \\
 \nonumber       + (1 - p_i)
     q_i\Big[p_jq_j((1-\psi_i)b +  \psi_i\tau + (1 - p_j)
      q_j(\psi_i(1 - \theta)\tau ) 
         p_j(1 - q_j)((1-\psi_i)b + \psi_i\theta \tau \Big] \\
   \nonumber     + p_i(1 - q_i)\Big[p_j
      q_j(1-\psi_i)(b - c/2) + (1 - p_j)
      q_j((1-\psi_i)(b - c) + \psi_i\theta \tau) \\
  \nonumber    + p_j(1 - q_j)((1-\psi_i)(b - c/2) + \psi_i(1 - \theta)\tau ) + (1 - 
        p_j)(1 - q_j)((1-\psi_i)(b - c) + \psi_i\tau )\Big] \\
    \nonumber    + (1 - 
       p_i)(1 - q_i)\Big[p_jq_j((1-\psi_i)b + \psi_i(1-\tau)) + (1 - p_j)
      q_j(\psi_i(1-\tau) + \psi_i\theta \tau)\\
      + p_j(1 - q_j)((1-\psi_i)b + \psi_i(1-\tau) + \psi_i(1 - \theta)\tau) + (1 - p_j)(1 - 
        q_j)(\psi_i(1-\tau) +  \psi_i\tau)\Big]
\end{align}
which is identical to Eq. 11, with $\psi=\psi_i$.

\subsubsection{Selection on $\psi$}
In order to identify the stable the equilibria of the three-trait model, we first look for evolutionary singular strategies with $\psi^*_r>0$, which must satisfy

\begin{eqnarray}
\nonumber  \psi_r^* &=&\frac{2 (b - c)  - p_r^*(2 b - c)}{2((1 - \tau) (1 - 2 q^*_r )+(b - c)-p_r^*(b - c/2))}\\
\nonumber p_r^*&=&1 - \sqrt{\frac{2 b - c-1}{2 b - c}}\\
\nonumber q_r^*&=&1-\frac{1}{2\tau(1-\theta)} - \frac{1-\tau(1-\theta)}{\tau(1-\theta)}p_r^*\\
\end{eqnarray}

The eigenvalues for $\mathbf{J}$ at this equilibrium are algebraically complex, however we show below that any equilibrium for which all three variables have zero selection gradient is disrupted by diversifying section, which leads to branching. 
\\
\\
Looking in detail at the selection gradient associated with $\psi_i$ we find

$$
\frac{\partial \pi_i}{\partial \psi_i}\Big |_{\psi_i=\psi_r}= 1 -(2 b - c) (2 - p_r^*) p_r^*-2 (1 - q_r^*) q_r^* - (p_r^* - q_r^*) (1 - 2 q_r^*) ((1-\tau) (1 - \theta) + \theta)
$$
which is independent of $\psi_r^*$. This yields three possible scenarios for the evolution of $\psi$ under low material payoffs.
\\
\\
\noindent \textbf{Case I: $\frac{\partial \pi_i}{\partial \psi_i}\Big |_{\psi_i=\psi_r}<0$.}
If $\frac{\partial \pi_i}{\partial \psi_i}\Big |_{\psi_i=\psi_r}<0$ at equilibrium, then the only viable equilibrium occurs at $\psi_r^*=0$. 

When $\psi=0$, the game reduces to the classic $2\times2$ snowdrift (when $b>c$) or a prisoner's dilemma (when $c/2<b<c$). The snowdrift game has a unique equilibrium at $p^*_r=\frac{2(b-c)}{2b-c}$. Note that in this scenario $q$ is released from selection (i.e. any choice of $q_r^*$ is an equilibrium). 

Whether this equilibrium is stable depends on $q_r^*$, as well as $b$, $c$, $\theta$ and $\tau$. However we note that the selection gradient depends on $b$ and $c$ through the term $ -(2 b - c) (2 - p_r^*) p_r^*$. Substituting $p^*_r=\frac{2(b-c)}{2b-c}$ this becomes
$-4\frac{b(1-c/b))}{2-c/b})$. Thus if we make $b\gg1$ while keeping $b/c$ fixed (i.e. keeping the $2\times2$ game of the same type), this term dominates all the other terms, and the equilibrium becomes stable. That is, we can always choose a large enough material payoff to make the equilibrium with $\psi_r^*=0$ stable.

When $2/c<b<c$ the game is a prisoner's dilemma with equilibrium $p_r^*=0$. In this case the selection gradient becomes 

$$
\frac{\partial \pi_i}{\partial \psi_i}\Big |_{\psi_i=\psi_r, p_r^*=0}= 1 + 2  \tau (q_r^*)^2 (1 - \theta) - q_r^* (1 +\tau(1-\theta))
$$
This is always positive, i.e. the equilibrium is unstable.
\\
\\
\noindent \textbf{Case II: $\frac{\partial \pi_i}{\partial \tau_i}\Big |_{\psi_i=\psi_r}>0$.}
If $\frac{\partial \pi_i}{\partial \psi_i}\Big |_{\psi_i=\psi_r}>0$ at equilibrium, then the only viable equilibrium occurs at $\psi_r^*=1$. 

We can use the stability conditions for the two-trait model, with $\psi=1$, to determine whether such an equilibrium can be stable. There are five possible cases as in the previous section 2.2.4. Taking in these in turn we find

\begin{enumerate}
\item Taking $q_r^*=0$ and $0<p_r^*<1$, and $\frac{\partial \pi_i}{\partial \psi_i}\Big |_{\psi_i=\psi_r}>0$. Applying the relevant stability condition (Case I from Section 2.1.3) we see that the equilibrium does not produce a viable $p_r^*$ unless $\tau=1$, which results in a stable equilibrium if $(2b-c)(2\theta-1)<c$ and is unstable otherwise. In this case the condition for $\frac{\partial \pi_i}{\partial \psi_i}\Big |_{\psi_i=\psi_r}>0$ is $-2 b^2 + 2 b (1 + c - \theta) + c ( 2 \theta-1) > 0$.
\item Taking $q_r^*=1$ and $0<p_r^*<1$, and $\frac{\partial \pi_i}{\partial \psi_i}\Big |_{\psi_i=\psi_r}>0$. Applying the relevant stability condition (Case II from Section 2.1.3) we see that the equilibrium does not produce a viable $p_r^*$ unless $\tau=1$, which results in a stable equilibrium if $c(2\theta-1)<2(b-c)$ and is unstable otherwise. In this case the condition for $\frac{\partial \pi_i}{\partial \psi_i}\Big |_{\psi_i=\psi_r}>0$ is $-2 b^2 + 2 b (1 + c) - c ( 1+\theta) > 0$
\item Taking $q_r^*=0$ and $p_r^*=0$, results in $\frac{\partial \pi_i}{\partial \psi_i}\Big |_{\psi_i=\psi_r}=1$ for all $\psi_r^*$ and is therefore always stable. 
\item Taking $q_r^*=1$ and $p_r^*=0$, results in $\frac{\partial \pi_i}{\partial \psi_i}\Big |_{\psi_i=\psi_r}=\tau(1-\theta)$, which is positive unless $\theta=1$. Applying the relevant stability condition (Case VII from Section 2.1.3) this is unstable unless $\tau=1$, in which case we require $-(b-c)\geq0$ and $\theta \leq 1/2$.
\item Taking $q_r^*=1$ and $p_r^*=1$, results in $\frac{\partial \pi_i}{\partial \psi_i}\Big |_{\psi_i=\psi_r}=\frac{1}{2}(2-2b+c)$ or all $\psi_r^*$, and is therefore either unconditionally stable or unconditionally unstable for a given choice of material payoffs, $b$ and $c$.
\end{enumerate}
Next we consider the case of zero selection gradient on $\psi$.
\\
\\
\noindent \textbf{Case III: $\frac{\partial \pi_i}{\partial \psi_i}\Big |_{\psi_i=\psi_r}=0$.}
Since there is no viable solution for $0<q_r^*<1$ when $\frac{\partial \pi_i}{\partial \psi_i}\Big |_{\psi_i=\psi_r}=0$, we look at the dynamics in the $p-\psi$ plane with $q=1$ and $q=0$. To do this we calculate the eigenvalues of $\mathbf{J}(p,\psi)$ where

$$
\mathbf{J}(p,\psi)=\begin{pmatrix}
\frac{\partial \mathbf{s}_p}{\partial p}  & \frac{\partial \mathbf{s}_p}{\partial \psi}\\
\frac{\partial \mathbf{s}_\psi}{\partial p}  & \frac{\partial \mathbf{s}_\psi}{\partial \psi}
\end{pmatrix}
$$

and $\mathbf{s}_\psi=\frac{\partial \pi_i}{\partial \psi_i}$ and $\mathbf{s}_p=\frac{\partial \pi_i}{\partial p_i}$ are the selection gradients of the two traits. The equilibrium of $\mathbf{J}$ have negative real part, and Eq. 12 is stable provided $\text{tr}(\mathbf{J})<0$ and $\text{det}(\mathbf{J})>0$. Calculating the trace and determinant of $\mathbf{J}$ at $(p_r^*,psi_r^*)$ we find 

\begin{eqnarray}
\nonumber    \text{tr}(\mathbf{J})&=& -\frac{1}{2} (1 -\psi) (2 b - c)\\
\nonumber     \text{det}(\mathbf{J})&=&-\Big[(2 b - c) (1 - p_r^*) + \theta\tau (1 - 2 q_r^*) + (1-\tau) (1 - 2 q_r^*)\Big]\Big[\frac{1}{2} (2 b - c) (1 - p_r^*) + (1-\tau) (1 - 2 q_r^*)\Big]
\end{eqnarray}

Since $2b>c$ is required for the game to be a Prisoner's Dilemma or a Snowdrift game, and the norm utility weights $\psi$, $\tau$ and $\theta$ all lie in $[0,1]$, $\text{det}(\mathbf{J})<0$ and the equilibrium is unstable. In addition, note that if $2b<c$, $\text{tr}(\mathbf{J})>0$ and the equilibrium is again unstable. 

Thus we conclude that no stable equilibrium exists in the dynamics of the $p-\psi$ plane. Finally we note that an equilibrium exists such that $\frac{\partial \pi_i}{\partial \psi_i}\Big |_{\psi_i=\psi_r}=0$ when $\psi_r^*=0$. In the next section we show that at this equilibrium, branching occurs.

\subsubsection{Branching in $p-q-\psi$}
We first look at the impact of diversifying selection on the equilibrium given in Eq. 17. 

We can  analyze the conditions for branching in the three-trait model by looking at the eigenvalues of

$$
\mathbf{H}(p,q,\psi)=\begin{pmatrix}
\frac{\partial^2 \pi_i}{\partial p_i^2}  & \frac{\partial^2 \pi_i}{\partial p_i\partial q_i} & \frac{\partial^2 \pi_i}{\partial p_i\partial \psi_i}\\
\frac{\partial^2 \pi_i}{\partial q_i\partial p_i}  & \frac{\partial^2 \pi_i}{\partial q_i^2} & \frac{\partial^2 \pi_i}{\partial q_i\partial \psi_i} \\
\frac{\partial^2 \pi_i}{\partial \psi_i\partial p_i}  & \frac{\partial^2 \pi_i}{\partial \psi_i\partial q_i} & \frac{\partial^2 \pi_i}{\partial \psi_i^2} 
\end{pmatrix}
$$

we find $\lambda_1=0$

$$
\lambda_{2/3}=\pm \sqrt{(-c (2 - p_r^*) + 2 b (1 - p_r^*) + 2 (1-\tau) (1 - 2 q_r^*))^2 + 4 (1-\tau)^2 4 \left(\psi_r^*\right)^2}
$$
and so there is always one negative and one positive eigenvalue, i.e. branching occurs.

In addition to the equilibrium given in Eq. 17, which holds for $\psi_r^*>0$, there is an additional equilibrium satisfying

\begin{eqnarray}
\nonumber  \psi_r^* &=&0\\
\nonumber p_r^*&=&\frac{2(b-c)}{2b-c}\\
\end{eqnarray}

such that $\frac{\partial \pi_i}{\partial \psi_i}\Big |_{\psi_i=\psi_r}=0$ (in contrast to the Case I above, for which it is assumed $\frac{\partial \pi_i}{\partial \psi_i}\Big |_{\psi_i=\psi_r}<0$). When $\psi_r^*=0$ the selection gradient on $q$ is zero for all values of $q_r^*$. Under the dynamics of our (simulation) model, this leads the value of $q_r^*$ to drift. When $q_r^*$ satisfies
$$
q_r^*=p_r^* + \frac{1 - 2 p_r^*}{2\tau(1 - \theta)}
$$
the selection gradient of all three traits is zero.



Calculating the eigenvalues of $\mathbf{H}(p,q,0)$ for $p_r^*=\frac{2(b-c)}{2b-c}$, $q_r^*=p_r^* + \frac{1 - 2 p_r^*}{2\tau(1 - \theta)}$ and $\psi_t^*=0$ we find $\lambda_1=0$

$$
\lambda_{2/3}=\pm \frac{(1-\tau) (2 b - 3 c) ((1-\tau)(1-\theta) + \theta)}{\tau (2 b + 
   c) (1 - \theta)}
$$
and so there is always one negative and one positive eigenvalue. Thus we expect evolutionary branching to occur at both equilibria. This is what we observe in simulations.

\subsubsection{Qualitative description of Nash equilibria}

Taken together, we find five qualitative types of equilibrium for the three-trait model as follows

\begin{enumerate}
    \item $p_r^*=0$, $q_r^*=0$, $\psi_r^*=0$ corresponding to high social homogeneity (no branching), high tightness, low cooperation. 
    \item $p_r^*=1$, $q_r^*=1$, $\psi=1$ corresponding to high social homogeneity (no branching), high tightness, high cooperation. At this equilibrium $\tau$ is released from selection.
    \item $p_r^*<1$, $q_r^*=1$, $\psi=1$ corresponding to high social homogeneity (no branching), low tightness, low cooperation.
    \item $p_r^*>0$, $q_r^*=0$, $\psi=1$ corresponding to high social homogeneity (no branching), low tightness, low cooperation.
    \item $p_r^*>0$, $q_r^*\geq 0$, $\psi_r^*=1$ corresponding to low social homogeneity (branching), low tightness, low cooperation. At this equilibrium $\psi$, $p$ and $q$ undergo branching.
\end{enumerate}

\subsubsection{Three-trait model with evolving $\theta$}

Next we consider a three-trait model, in which an individual $i$ is characterized by their probability of cooperation, $p_i$, their probability of expressing belief in cooperation, $q_i$ and their descriptive norm weight, $\theta_i$. We set all other parameters, $\tau$, $\psi$, $b$ and $c$ to be constant and the same for all players. 

The utility for a player $i$ interacting with a player $j$ under this model is

\begin{align}
\nonumber \pi_{ij}=p_iq_i
\Big[p_jq_j((1-\psi)(b - c/2) + \psi) + (1 - p_j)
      q_j((1-\psi)(b - c) + \psi(1-\tau) + (1 - \theta_i)\psi\tau ) \\
\nonumber      + p_j(1 - q_j)((1-\psi)(b - c/2) + \psi(1-\tau) + \psi\theta_i \tau ) + (1 - 
        p_j)(1 - q_j)((1-\psi)(b - c) + \psi(1-\tau))\Big] \\
 \nonumber       + (1 - p_i)
     q_i\Big[p_jq_j((1-\psi)b +  \psi\tau + (1 - p_j)
      q_j(\psi(1 - \theta_i)\tau ) 
         p_j(1 - q_j)((1-\psi)b + \psi\theta_i \tau \Big] \\
   \nonumber     + p_i(1 - q_i)\Big[p_j
      q_j(1-\psi)(b - c/2) + (1 - p_j)
      q_j((1-\psi)(b - c) + \psi\theta_i \tau) \\
  \nonumber    + p_j(1 - q_j)((1-\psi)(b - c/2) + \psi(1 - \theta_i)\tau ) + (1 - 
        p_j)(1 - q_j)((1-\psi)(b - c) + \psi\tau )\Big] \\
    \nonumber    + (1 - 
       p_i)(1 - q_i)\Big[p_jq_j((1-\psi)b + \psi(1-\tau)) + (1 - p_j)
      q_j(\psi(1-\tau) + \psi\theta_i \tau)\\
      + p_j(1 - q_j)((1-\psi)b + \psi(1-\tau) + \psi(1 - \theta_i)\tau) + (1 - p_j)(1 - 
        q_j)(\psi(1-\tau) +  \psi\tau)\Big]
\end{align}
which is identical to Eq. 11, with $\theta=\theta_i$.

\subsubsection{Selection on $\theta$}

In order to identify the stable equilibria of the three-trait model with evolving $\theta$, we first look for evolutionary singular strategies, which satisfy

$$
\frac{\partial \pi_i}{\partial q_i}\Big |_{q_i=q_r}=\frac{\partial \pi_i}{\partial p_i}\Big |_{p_i=p_r}=\frac{\partial \pi_i}{\partial \theta_i}\Big |_{\theta_i=\theta_r}=0
$$
This yields a single solution

\begin{eqnarray}
\nonumber q_r^*&=&\frac{1}{2}\\
\nonumber p_r^*&=&\frac{2 (b - c) }{2 b - c} \\
\nonumber \theta_r^*&=&-\frac{1-\tau}{\tau}\\
\end{eqnarray}

This only produces a physical solution in the limiting case $\tau=1$ (preference for conformity), in which case $\theta_r^*=0$. 

Looking in detail at the selection gradient associated with $\theta_i$ we find

$$
\frac{\partial \pi_i}{\partial \theta_i}\Big |_{\theta_i=\theta_r}=-\psi (p_r^* - q_r^*) (1 - 2 q_r^*) \tau
$$
which is of the same form as the selection gradient on $\tau$. The same analysis can be performed as in Section 2.2.2. The result is that equilibria emerge in which either $\theta_r^*=0$ (injunctive norms) or $\theta$ is released from selection (if $\tau=0$, $\psi=0$ or $p_r^*=q_r^*$).

We next look at the conditions for branching when $\tau=0$ and $\psi>0$. Calculating the eigenvalues for $\mathbf{H}(p,\theta)$ where

$$
\mathbf{H}(p,\theta)=\begin{pmatrix}
\frac{\partial^2 \pi_i}{\partial p_i^2}  & \frac{\partial^2 \pi_i}{\partial p_i\partial \theta_i}\\
\frac{\partial^2 \pi_i}{\partial \theta_i\partial p_i}  & \frac{\partial^2 \pi_i}{\partial \theta_i^2}
\end{pmatrix}
$$
we find $\lambda=0$ and so diversifying selection does not promote branching when $\theta$ can evolve in the three-trait model.

\subsection{Five-trait model}

Finally we give a brief discussion of the five-trait model in which $p$, $q$, $\tau$, $\psi$ and $\theta$ evolve. In particular we note that the only viable equilibria for which multiple traits have zero selection gradient occur when

\begin{eqnarray}
\nonumber  \psi_r^* &=&0\\
\nonumber p_r^*&=&\frac{2(b-c)}{2b-c}
\end{eqnarray}
Which gives us the same scenario as discussed for branching in the three-trait model with evolving $\psi$ above (section 2.2.7).

We can  analyze the conditions for branching in the three-trait model by looking at the eigenvalues of

$$
\mathbf{H}(p,q,\psi,\tau,\theta)=\begin{pmatrix}
\frac{\partial^2 \pi_i}{\partial p_i^2}  & \frac{\partial^2 \pi_i}{\partial p_i\partial q_i} & \frac{\partial^2 \pi_i}{\partial p_i\partial \psi_i} & \frac{\partial^2 \pi_i}{\partial p_i\partial \tau_i} & \frac{\partial^2 \pi_i}{\partial p_i\partial \theta_i}\\
\frac{\partial^2 \pi_i}{\partial q_i\partial p_i} & \frac{\partial^2 \pi_i}{\partial q_i^2}  &  \frac{\partial^2 \pi_i}{\partial q_i\partial \psi_i} & \frac{\partial^2 \pi_i}{\partial q_i\partial \tau_i} & \frac{\partial^2 \pi_i}{\partial q_i\partial \theta_i}\\
\frac{\partial^2 \pi_i}{\partial \psi_i\partial p_i} & \frac{\partial^2 \pi_i}{\partial \psi_i\partial q_i}  &  \frac{\partial^2 \pi_i}{\partial \psi_i^2} & \frac{\partial^2 \pi_i}{\partial \psi_i\partial \tau_i} & \frac{\partial^2 \pi_i}{\partial \psi_i\partial \theta_i}\\
\frac{\partial^2 \pi_i}{\partial \tau_i\partial p_i} & \frac{\partial^2 \pi_i}{\partial \tau_i\partial q_i}  &  \frac{\partial^2 \pi_i}{\partial \tau_i\partial \psi_i}  & \frac{\partial^2 \pi_i}{\partial \tau_i^2} & \frac{\partial^2 \pi_i}{\partial \tau_i\partial \theta_i}\\
\frac{\partial^2 \pi_i}{\partial \theta_i\partial p_i} & \frac{\partial^2 \pi_i}{\partial \theta_i\partial q_i}  &  \frac{\partial^2 \pi_i}{\partial \theta_i\partial \psi_i}  & \frac{\partial^2 \pi_i}{\partial \theta_i\partial \tau_i} & \frac{\partial^2 \pi_i}{\partial \theta_i^2}
\end{pmatrix}
$$

Calculating the eigenvalues of $\mathbf{H}(p,q,\psi,\tau,\theta)$ for $p_r^*=\frac{2(b-c)}{2b-c}$ and $\psi_r^*=0$ we find non-zero eigenvalues

$$
\lambda=\pm \sqrt{h(b,c,q_r^*,\tau_r^*,\theta_r^*)}
$$
The precise form of the function $h$ is algebraically complicated and can be found in the Mathematica notebook (link to github). However we note that, numerically, we find viable parameter values such that $h>0$ indicating the presence of branching in the five-trait model (which indeed we observe in our simulations as seen in main text Figure 3). Further details of the equilibria in the five-trait model are shown in Figure S7-S8

\clearpage

\section{Additional Simulations}

Here we present additional simulation results, focused in particular on the conditions for evolutionary branching to occur.

\subsection{Classifying branched simulations}   

In addition to the analytic conditions for branching, we also identify instances of branching in our simulations. We use a highly conservative definition of branching to classify the output of our simulations. We first note that branching always occurs in $p$ (along with at least one other variable). We next note that there exists a pure strategy Nash equilibrium in which some players always defect and others always cooperate. Finally we observe in our simulations that branching always results in one group with $p=0$ (always defect) and another with always cooperate $p=1$. 

Thus we classify a simulation as "branched" if i) the population contains both individuals with $p<0.05$ and $p>0.95$ and ii) that the population remains in this state for at lease 1000 generations (i.e. $1000N$ updates of the imitation dynamics). Figure S4 shows an example of the kind of output observed. Here we see branching in $p-\tau$ (three trait model with evolving $\tau$). We see the population move to and remain at a state in which one group cooperates and the other defects.

\begin{figure*}[!h]
\centering
\includegraphics[width=0.8\textwidth]{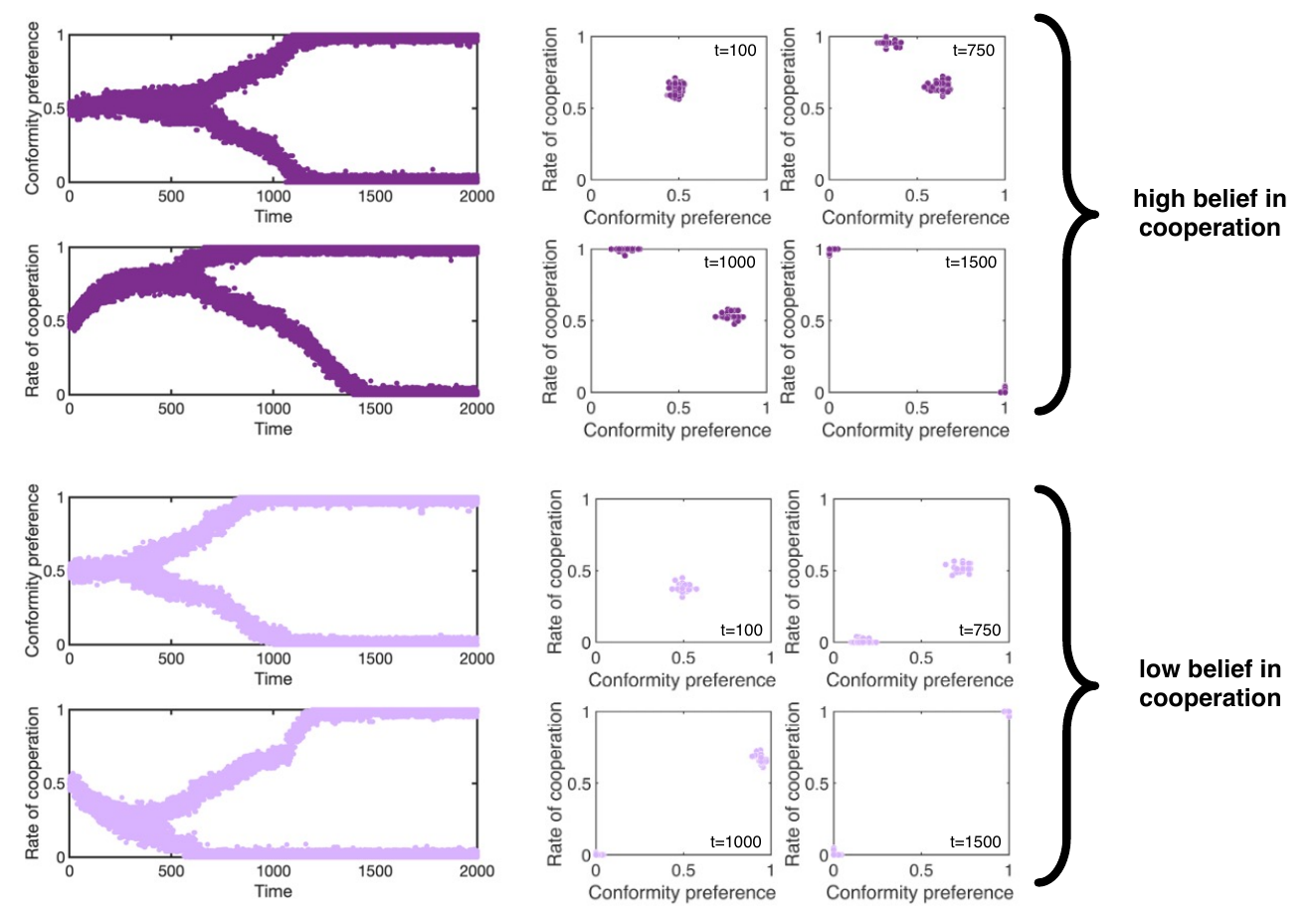}
\caption*{\textbf{Figure S4 - After branching the groups evolve to either $p=1$ or $p=0$.} }\end{figure*}

\subsection{Varying material payoffs}

Next we show the detailed effects of varying the cost of cooperation on the rate of cooperation and the conformity preference $\tau$ for each of the six equilibria of the system (see main text Figure 2).

\begin{figure*}[!h]
\centering
\includegraphics[width=1\textwidth]{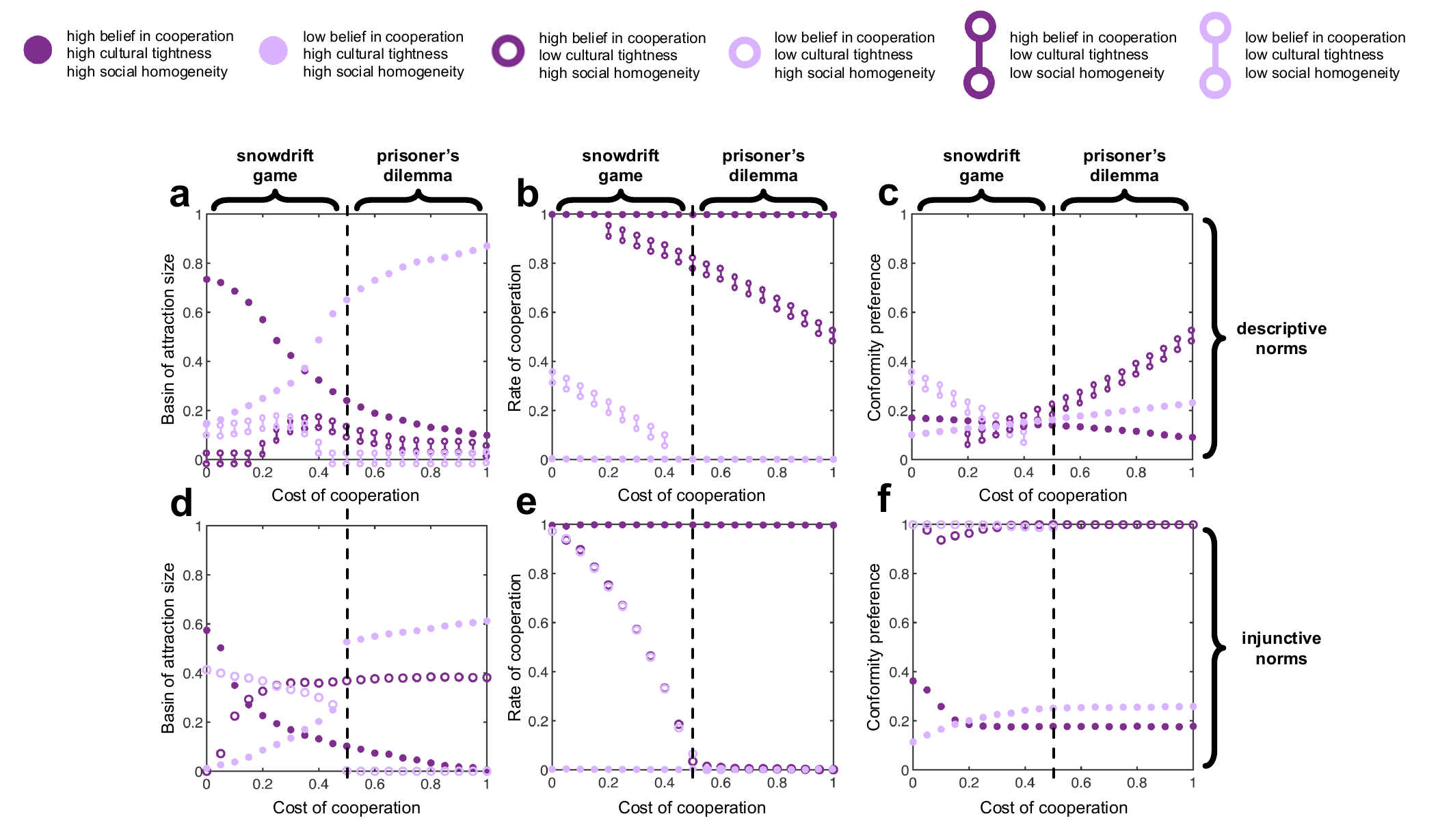}
\caption*{\textbf{Figure S5 - Impact of varying cost of cooperation}, $c$ on the basin of attraction (a,d), rate of cooperation (b,e) and conformity preference, $\tau$ (c,f) under both descriptive (top row, $\theta=1$) and injunctive norms (bottom row, $\theta=0$(. These results  are used to produce the ``cartoon'' in Figure 2a. Simulations and parameter choices are the same as described for Figure 2. }\end{figure*}

\subsection{Varying descriptive norm weight, $\theta$}

Next we consider the effect of systematically varying the descriptive norm weight $\theta$ on the outcome of the three-trait model with evolving $\tau$ (Figure S6).

\begin{figure*}[!h]
\centering
\includegraphics[width=1\textwidth]{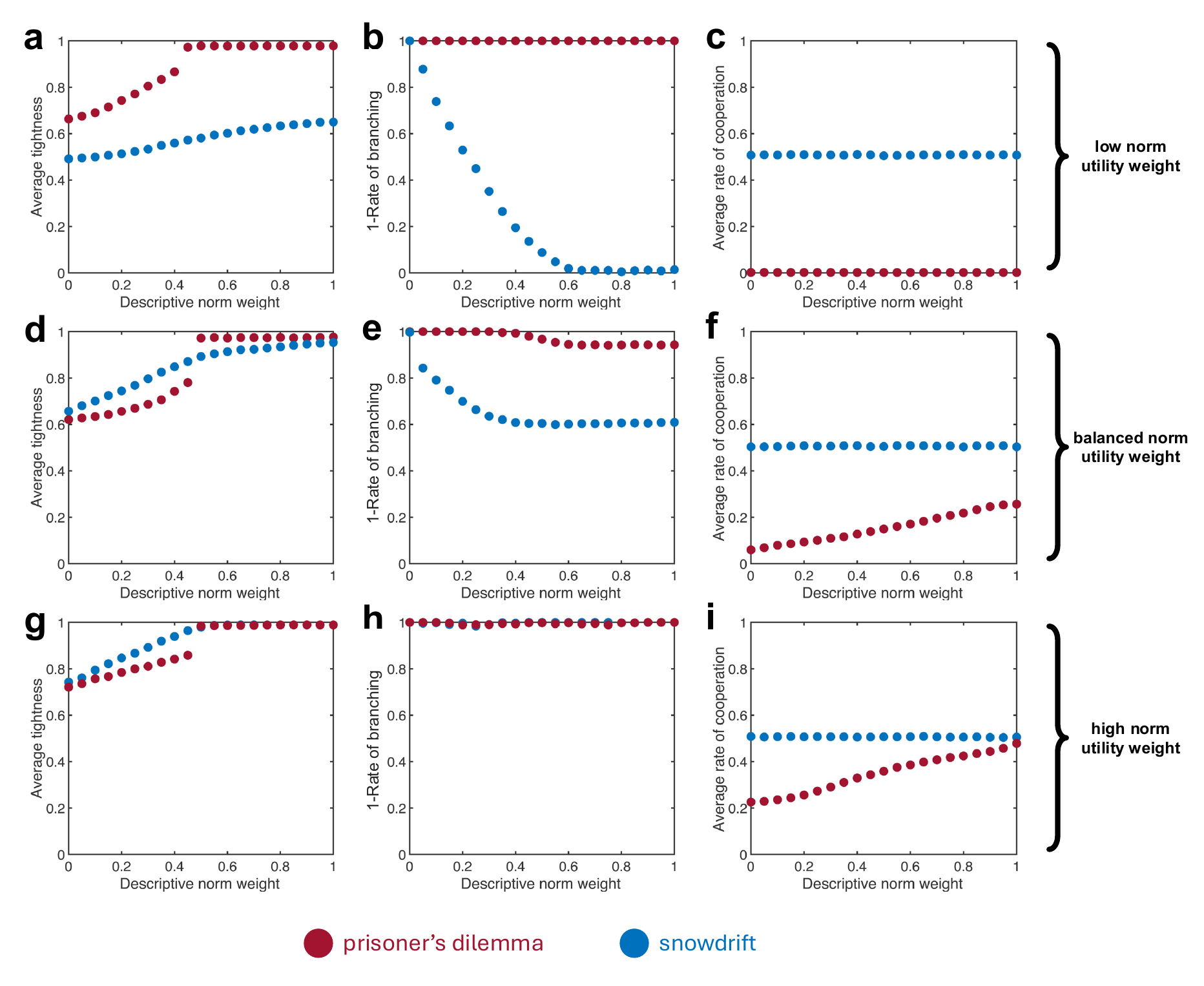}
\caption*{\textbf{Figure S6 - Effect of varying the descriptive norm weight, $\theta$ on the behavior of the three-trait model with evolving $\tau$}. Results are shown for material payoffs corresponding to a prisoner's dilemma (red, $c=1$, $b=2/3$) and for the snowdrift game (blue, $c=1$, $b=3/2$). In the middle row we show $1-\text{rate of branching}$ for different norm utility weights $\psi=0.1$ (low, top), $\psi=0.5$ (balanced, middle) and $\psi=0.9$ (high, bottom). We see that when norm utility rate is low, branching in the snowdrift game is almost certain once $\theta$ becomes large enough. For intermediate norm utility weights branching remains likely once $\theta$ becomes large enough, and can also occur (rarely) in the prisoner's dilemma. Once norm utility weight becomes high, branching does not occur. }\end{figure*}

We note in particular that, for intermediate norm utility weights, branching is observed at low frequency in the prisoner's dilemma.

\clearpage

\subsection{Five-trait model}

In the five-trait model we observe branching in $\psi-p-q$ similar to the three-trait model with evolving $\psi$ (main text Figure 3). Figure S7 shows how the basin of attraction of different equilibria vary with the benefit of cooperation $b$ (keeping $b/c$ fixed).

\begin{figure*}[!h]
\centering
\includegraphics[width=0.8\textwidth]{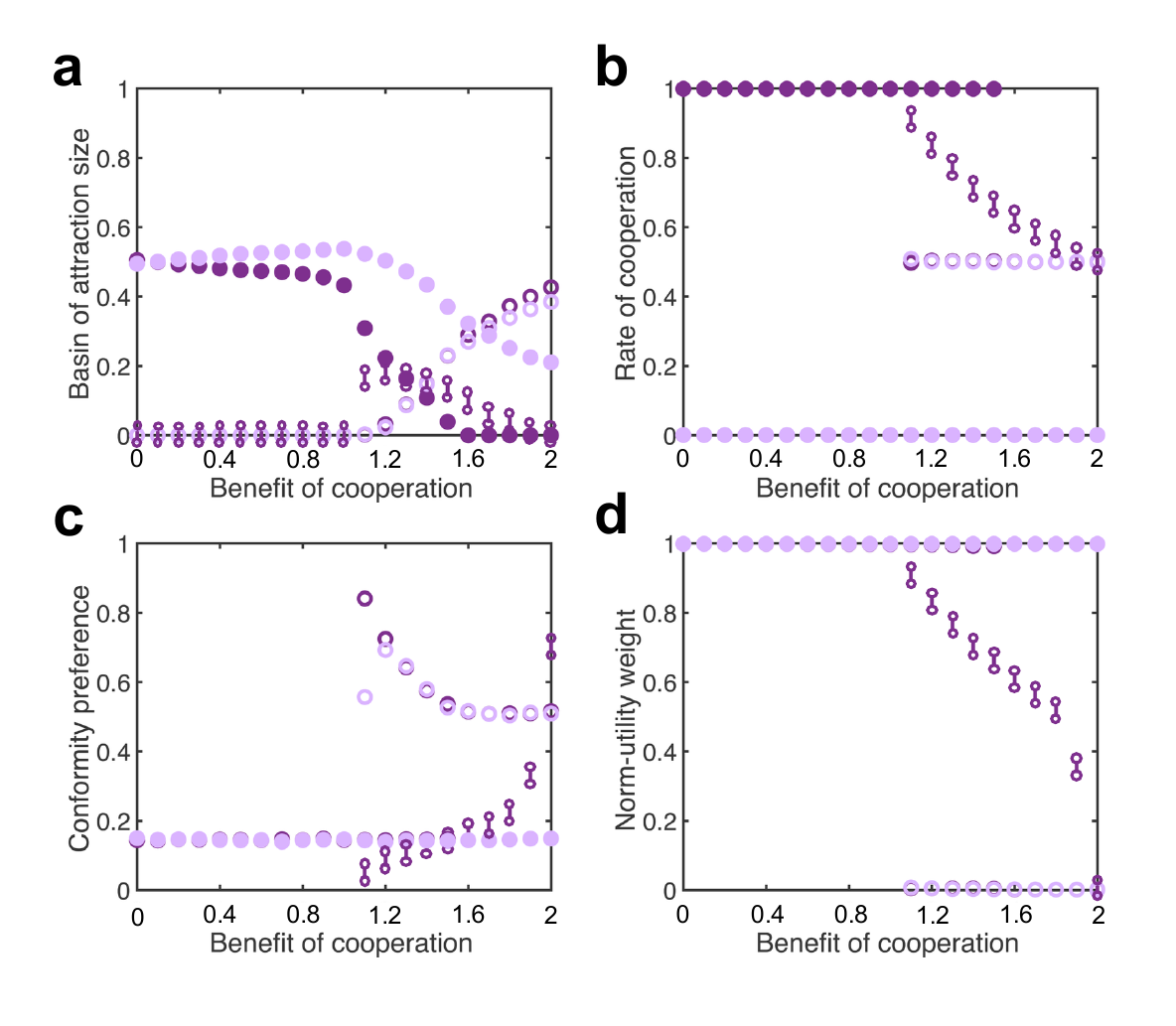}
\caption*{\textbf{Figure S7 - Basins of attraction in the five-trait model.} We observe the same equilibria as for the three-trait model with evolving $\psi$ (main text Figure 3). a) We vary the benefit of cooperation $b$, which keeping $b/c=3/2$ to ensure a snowdrift game. Once the benefit of cooperation becomes big enough, a branched equilibrium emerges. b) Rate of cooperation associated with each equilibrium. c) Evolved conformity preference for each equilibrium. d) Evolved norm-utility weight associated with each equilibrium. Simulations use the same parameters as described in main text Figure 3, with all five parameters $\theta$, $\psi$, $\tau$, $p$ and $q$ allowed to evolve.}\end{figure*}

Figure S8 shows a specific instance of branching in the three trait model. We see that one group emerges with $\psi=0$, which results in the parameters $q$, $\theta$ and $\tau$ being released from selection.

\begin{figure*}[!h]
\centering
\includegraphics[width=0.8\textwidth]{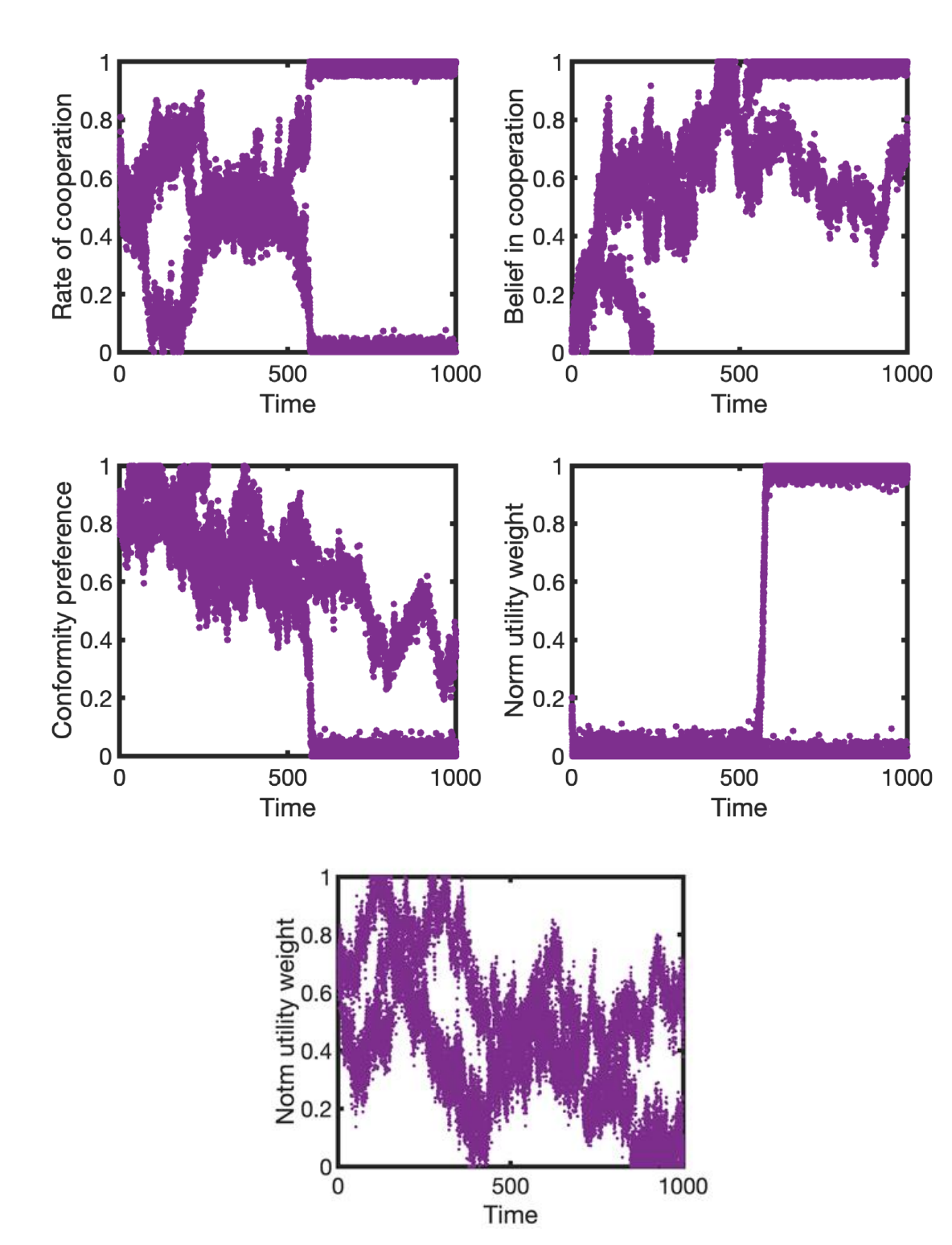}
\caption*{\textbf{Figure S8 - Branching in the five trait model}. Shown is a specific instance of branching in the five trait model, with $N=100$, $b=1.2$ and $b/c=3/2$, with simulations as described for main text Figures 2-3. We see that one group, which evolves to $\psi=0$, experiences ``drift'' in $q$, $\theta$ and $\tau$.}\end{figure*}

\clearpage


\end{document}